\documentclass{emulateapj}
\newlength{\bigfigsize}
\newif\iftwocol\twocoltrue

\usepackage{ifthen}
\newif\ifallcolourfigs\allcolourfigstrue
\newif\ifincludethreed\includethreedfalse
\newif\ifincludehighres\includehighresfalse

\newif\ifpersonal\personalfalse
\newif\ifarxiv\arxivtrue
\newif\ifapj\apjfalse

\newcommand{\figuredir}{.}
\ifpersonal
	\allcolourfigstrue
	\includethreedfalse
	\includehighresfalse
	\newcommand{\choosefigure}[3]{\figuredir/#1}
	\newcommand{\choosefigurea}[3]{\figuredir/#1}
	\newcommand{\choosefigureb}[3]{\figuredir/#1}
	
\fi
\ifarxiv
	\allcolourfigstrue
	\includethreedfalse
	\includehighresfalse
	\newcommand{\choosefigure}[3]{\figuredir/#2}
	\newcommand{\choosefigurea}[3]{\figuredir/#2}
	\newcommand{\choosefigureb}[3]{\figuredir/#2}
	
\fi
\ifapj
	\allcolourfigstrue
	\includethreedfalse
	\includehighresfalse
	\newcommand{\choosefigure}[3]{\figuredir/#3}
	\newcommand{\choosefigurea}[3]{\figuredir/#3}
	\newcommand{\choosefigureb}[3]{\figuredir/#3}
	
\fi

\usepackage{graphicx,color}
\usepackage{natbib}
\usepackage{amssymb}
\usepackage{apjfonts}

\newcommand{\paperurl}{{http://www.cita.utoronto.ca/$\sim$ljdursi/draping/}}
\newcommand{\forthreedfiguressee}{\ifincludethreed {} \else A PDF version of this manuscript with an interactive 3d version of this figure, following \cite{3dinpdf}, is available at \paperurl .\fi}
\newcommand{\forhighresfiguressee}{\ifincludehighres {} \else This figure has been somewhat degraded.   For a PDF version of this manuscript with higher resolution figures and interactive 3d graphics, please see \paperurl .\fi}

\newcommand{\del}[1]{{}}

\newcommand{\MachAlfven}{\ensuremath{{\cal{M}}_A}}
\newcommand{\partf}[2]{\ensuremath{\frac{\partial #1}{\partial #2}}}

\newcommand{\FLASH}{{\sc{Flash}}{}}
\newcommand{\eg}{{\it{e.g.}}}
\newcommand\citeeg[1]{\citep[\eg{},][]{#1}}
\newcommand\citenote[2]{\citep[#1][]{#2}}

\newcommand{\bra}{\langle}
\newcommand{\ket}{\rangle}

\newcommand{\mathbfit}[1]{\textbf{\textit{#1}}}

\newcommand{\dd}{\mathrm{d}}
\newcommand{\rmn}{\mathrm}
\newcommand{\dps}{\displaystyle}
\newcommand{\eps}{\varepsilon}
\newcommand{\vecbf}{\mathbfit}
\newcommand{\B}{\vecbf{B}}
\newcommand{\x}{\vecbf{x}}
\newcommand{\e}{\vecbf{e}}
\renewcommand{\u}{\vecbf{u}}
\renewcommand{\b}{\vecbf{b}}
\renewcommand{\j}{\vecbf{j}}
\renewcommand{\v}{\upsilon}
\newcommand{\vel}{\bupsilon}
\newcommand{\Alfvenvel}{{\bupsilon_A}}
\newcommand{\Alfvenvelmag}{{\upsilon_A}}
\newcommand{\Reynolds}{\mathrm{Re}}

\SetSymbolFont{symbols}{bold}{OMS}{cmsy}{b}{n}
\DeclareSymbolFont{bmisymbols}{OML}{cmm}{b}{it}
\DeclareMathSymbol{\balpha}{0}{bmisymbols}{"0B}
\DeclareMathSymbol{\bbeta}{0}{bmisymbols}{"0C}
\DeclareMathSymbol{\bgamma}{0}{bmisymbols}{"0D}
\DeclareMathSymbol{\bdelta}{0}{bmisymbols}{"0E}
\DeclareMathSymbol{\bepsilon}{0}{bmisymbols}{"0F}
\DeclareMathSymbol{\bzeta}{0}{bmisymbols}{"10}
\DeclareMathSymbol{\boldeta}{0}{bmisymbols}{"11}
\DeclareMathSymbol{\btheta}{0}{bmisymbols}{"12}
\DeclareMathSymbol{\biota}{0}{bmisymbols}{"13}
\DeclareMathSymbol{\bkappa}{0}{bmisymbols}{"14}
\DeclareMathSymbol{\blambda}{0}{bmisymbols}{"15}
\DeclareMathSymbol{\bmu}{0}{bmisymbols}{"16}
\DeclareMathSymbol{\bnu}{0}{bmisymbols}{"17}
\DeclareMathSymbol{\bxi}{0}{bmisymbols}{"18}
\DeclareMathSymbol{\bpi}{0}{bmisymbols}{"19}
\DeclareMathSymbol{\brho}{0}{bmisymbols}{"1A}
\DeclareMathSymbol{\bsigma}{0}{bmisymbols}{"1B}
\DeclareMathSymbol{\btau}{0}{bmisymbols}{"1C}
\DeclareMathSymbol{\bupsilon}{0}{bmisymbols}{"1D}
\DeclareMathSymbol{\bphi}{0}{bmisymbols}{"1E}
\DeclareMathSymbol{\bchi}{0}{bmisymbols}{"1F}
\DeclareMathSymbol{\bpsi}{0}{bmisymbols}{"20}
\DeclareMathSymbol{\bomega}{0}{bmisymbols}{"21}
\DeclareMathSymbol{\bvarepsilon}{0}{bmisymbols}{"22}
\DeclareMathSymbol{\bvartheta}{0}{bmisymbols}{"23}
\DeclareMathSymbol{\bvarpi}{0}{bmisymbols}{"24}
\DeclareMathSymbol{\bvarrho}{0}{bmisymbols}{"25}
\DeclareMathSymbol{\bvarsigma}{0}{bmisymbols}{"26}
\DeclareMathSymbol{\bvarphi}{0}{bmisymbols}{"27}

\begin{document}

\title{Draping of Cluster Magnetic Fields over Bullets and Bubbles -- Morphology and Dynamic Effects}
\shorttitle{Magnetic Draping in Clusters}
\shortauthors{Dursi \& Pfrommer}

\author{L. J. Dursi and C. Pfrommer}
\affil{Canadian Institute for Theoretical Astrophysics,
                 University of Toronto, 
                 Toronto, ON, M5S~3H8, 
                 Canada}
\email{ljdursi@cita.utoronto.ca, pfrommer@cita.utoronto.ca}

\begin{abstract}
High-resolution X-ray observations have revealed cavities and `cold
fronts' with sharp edges in temperature and density within galaxy
clusters.  Their presence poses a puzzle since these features are
not expected to be hydrodynamically stable, or to remain sharp in
the presence of diffusion.  However, a moving core or bubble in
even a very weakly magnetized plasma necessarily sweeps up enough
magnetic  field to build up a dynamically important sheath; the
layer's strength is set by a competition between `plowing up' and
slipping around of field lines, and depends primarily on the ram
pressure seen by the moving object.  In this inherently three
dimensional problem, our analytic arguments and numerical experiments
show that this layer modifies the dynamics of a plunging core,
greatly modifying the hydrodynamic instabilities and mixing, changing
the geometry of stripped material, and slowing the core through
magnetic tension.  We derive an expression for the maximum magnetic
field strength and thickness of the layer, as well as for the
opening angle of the magnetic wake. The morphology of the magnetic
draping layer implies the suppression of thermal conduction across
the layer, thus conserving strong temperature gradients.  The
intermittent amplification of the magnetic field as well as the
injection of MHD turbulence in the wake of the core is identified
to be due to vorticity generation within the  magnetic draping
layer. These results have important consequences for understanding
the complex gasdynamical processes of the intra-cluster medium, and
apply quite generally to motions through other magnetized environments,
\eg{} the ISM.
\end{abstract}


\section{INTRODUCTION}
\label{sec:intro}
Recent observations of very sharp `cold fronts' in galaxy clusters
raise unanswered questions in the hydrodynamics of galaxy clusters
\citenote{see for instance the review of}{ShocksAndColdFrontsReview},
for such abrupt transitions are not expected to be stable against either
hydrodynamical motions or diffusion for extended periods of time.

\begin{figure}[h!]
\centering
\includegraphics[angle=-90,width=\bigfigsize]{\choosefigure{bubblewake-tubes-mag-morelines-normalvel.eps.epsf}{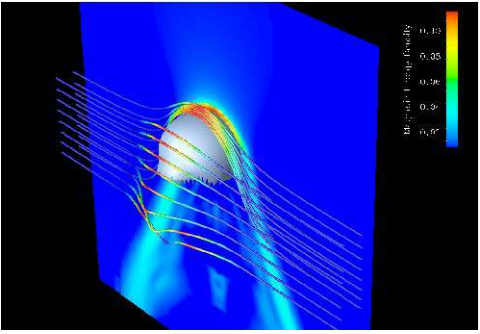}{bubblewake-tubes-mag-morelines-normalvel-cmyk.eps}}
\caption{A rendering of one of the three-dimensional simulations (referred to as Run F later in
this work) performed for this work, discussed in more detail in later sections but included here
to illustrate the physical picture.   An overdense projectile is sent through
a uniformly magnetized medium, sweeping up magnetic field ahead of it.
Plotted is a density isosurface, corresponding to the mean density of the
bullet, and some fiducial magnetic field lines.   The cut-plane is coloured
by magnetic energy density, as are the field lines.  The minimum of the colourmap for the
magnetic energy density is set to be the magnetic energy density in the ambient medium;
energies less than that (which occur in the wake, due to entrainment of unmagnetized material as the bubble rises)
is set to the color of the ambient energy density.  The magnetic field is
`draped' into a thin layer forming a bow wave, leaving turbulence in a
wake behind the bullet.   Magnetic field lines pile up along the stagnation
line of this initially axisymmetric bullet, while in the plane perpendicular
to the initial field, the field lines can slip around the bullet.   \forthreedfiguressee}
\label{fig:3drendering-normalvel}
\end{figure}

It has been known for some decades in the space science community that an
object moving super-Alfv\'enically in a magnetized medium can very rapidly
sweep up a significant magnetic layer which is then `draped' over the
projectile \citeeg{1980Ge&Ae..19..671B}.  For concreteness in discussing the
process, we show in Fig.~\ref{fig:3drendering-normalvel} a picture of this
mechanism taken from one of our simulations, which will be described in more
detail in later sections.

There has been significant interest in applying this idea of magnetic
draping in galaxy clusters \citep[\eg{},][]{vikhlininetal2001,
lyutikovdraping, asai04, asai05, asai07} as such a magnetic
field could naturally inhibit thermal
conduction across a front \citeeg{ettorifabian} allowing it to remain
sharp over dynamically long times.  Although such draping has been
explored in the past, in the space sciences the resulting dynamics is
relatively simpler, as generally the object being draped is a solid
body, with little interior dynamics of its own.  However, in the case
of for instance a merger of gas-rich clusters, the hydrodynamics of the
draped plunging core can also be modified, with the strong magnetic layer
providing some stabilization against instabilities that would otherwise
occur \citep{lintheory}.

The effect of a strong draped magnetic layer could be even greater for
underdense objects, such as for bubbles moving through the intercluster medium,
as seen at the centers of many cool-core clusters \citeeg{2005Natur.433...45M,
  2004ApJ...607..800B}.  In this case, the bubble would be quickly disrupted on
rising absent some sort of support \citeeg{flashbubbles}.  However, the draping
of a pre-existing magnetic field may strongly alter the dynamics and suppress
hydrodynamic instabilities, as seen recently in simulations
\citep{ruszkowskiI}. The morphology of the draped magnetic field may be able to
suppress transport processes across the bubble interface such as cosmic ray
diffusion and heat conduction. This has important consequences for cosmic ray
confinement in these buoyantly rising bubbles and may explain re-energized
radio `relic' sources, broad central abundance profiles of clusters, and the
excitation of the H$\alpha$ line in filaments trailing behind bubbles
\citep{ruszkowskiII}.  Although the analytics and simulations we discuss here
focus on the case of an overdense `core' moving in an external field, we expect
the basic magnetic dynamics to also extend to the case of an underdense bubble
probably depending on the magnetic energy density of the plasma.

Because we are interested here in the fundamentals of a basic process
--- that of the draping of a field around an object and the resulting
hydrodynamical effect on the object and its interaction with the
external medium ---  we consider for this paper, in both our analytic
and computational work, the simplest possible case; an overdense,
non-self-gravitating `blob' moving through a quiescent medium with a
magnetic field uniform on the scales considered.  (The term we will use
for this blob will depend on the situation; when discussing astrophysical
implications, we will speak of `cores' or `bullets', depending on the
circumstances, or `bubbles' for underdense regions; for the case of our
numerical simulations, we will refer to `projectiles', as the overdense
fluid in the simulations differs in structure from cores or bullets in
lacking self gravity; in our analytic work where the blob is a rigid
sphere, we will refer to the blob as a sphere or spherical body.)

We further consider the case of the object moving subsonically; while the case
of supersonic motion is interesting and highly relevant, we anticipate that in
the usual case where the bow shock is well separated from the magnetic layer --
that is where the standoff distance $\Delta \approx \frac{\rho_0}{\rho_s} R
\sim R$, (where $R$ is the radius of the core, $\rho_0$ is the ambient density,
and $\rho_s$ is the shocked density) is much greater than the magnetic layer
thickness $l \approx {\cal{M}}_A^{-2} R \ll R$, (where $l$ is the approximate
magnetized layer thickness and ${\cal{M}}_A$ the alfv{\'e}nic Mach number, as
discussed in more detail in the next section) that the arguments here will also
hold, so we save the more complicated geometry and larger parameter space of
the compressible case for future work.

\ifincludethreed
\begin{figure}[h!]
\centering
\includemovie[poster,3Dcoo=79.5 75 79.5, 3Droo=434.3960784034515, 3Dc2c=0 1 0]{5in}{5in}{3d/normalvel-small-compressed.u3d}
\caption{Interactive 3D version of Figure \ref{fig:3drendering-normalvel} above, following \cite{3dinpdf}.}
\end{figure}
\fi

In \S\ref{sec:picture} we give an overview of the physics of draping, putting
our work in the context of previous results; in \S\ref{sec:methodology} we
describe our analytic and computational approaches; in \S\ref{sec:comparison}
we compare the results of our two approaches, and from the understanding
gained there in \S\ref{sec:characteristics} we describe characteristics of
draping; we discuss the effect of instabilities in \S\ref{sec:instabilities},
consider the limitations of our results and consider applying them to later
times in \S\ref{sec:discussion}, and finally conclude in \S\ref{sec:conclusion}.

\section{GENERAL PHYSICS OF MAGNETIC DRAPING}
\label{sec:picture}
Previous work \citeeg{1980Ge&Ae..19..671B,lyutikovdraping} has looked at the
basic picture of magnetic draping in a simplified way in some detail; we
summarize some of their key results as well as our new insight into this
problem here.  In these works, the known potential flow around a solid sphere
is taken as an input, and a purely kinematic magnetic field, uniform and
perpendicular to the direction of motion, is added.  The derivation of
\cite{1980Ge&Ae..19..671B} is clarified, and a novel set of useful
approximations for the resulting field near the solid sphere are given, in
Appendix~\ref{sec:analytics}.

Because in this case the flow falls quickly to zero at the surface
of the moving sphere, magnetic field rapidly `builds up' around the projectile,
and in the kinematic limit eventually becomes infinite.
The high degree of symmetry along the stagnation line (the axis of
symmetry of the object pointing in the direction of motion)  greatly
simplifies the mathematics, and as shown in for instance
\cite{lyutikovdraping}, the magnetic field strength directly along
the stagnation line is given by
\begin{equation}
\frac{|B|}{\rho} = \frac{B_0}{\rho_0} \frac{1}{\sqrt{1 - \left(\frac{R}{R + s}\right)^3}}
\label{eq:stagline-bfield}
\end{equation}
where $B_0$ is the ambient magnetic field, $\rho_0$ is the ambient
density, $R$ is the radius of the solid sphere projectile, and $s$
is the distance along the stagnation line from the surface of the
sphere.

\begin{figure}[h!]
\centering
\plotone{\choosefigure{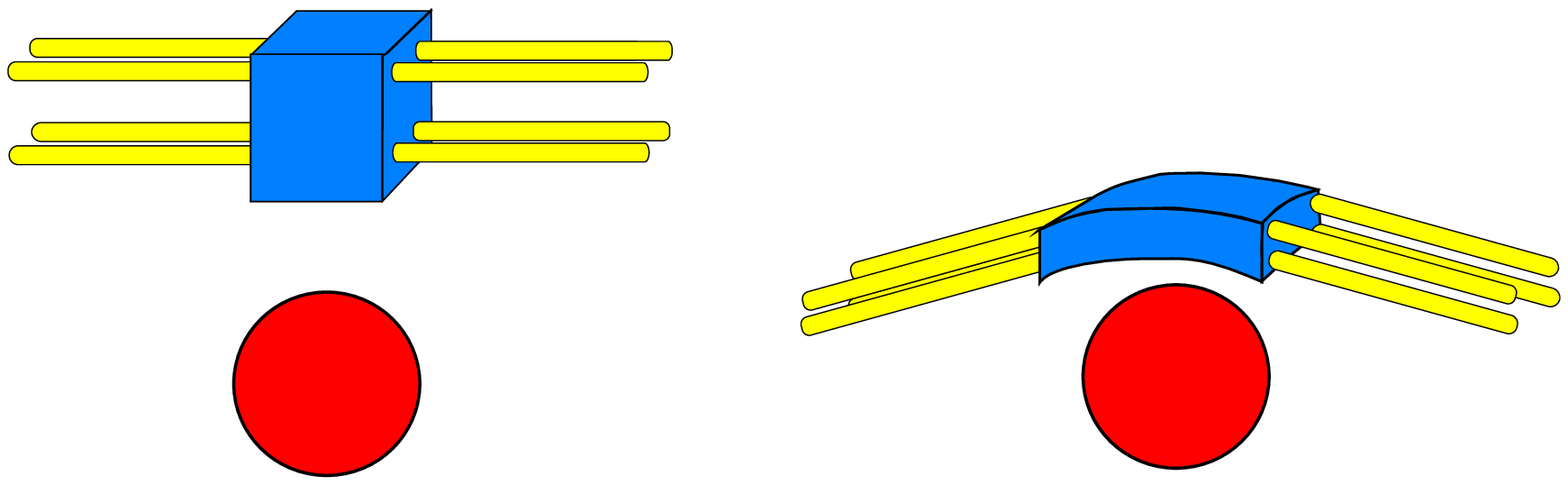} {draping.eps} {draping-bw.eps}}
\caption{A cartoon showing the distortion of incoming fluid elements 
and stretching of field lines as a red spherical projectile moves upwards
through the ambient medium.}
\label{fig:draping-cartoon}
\end{figure}

The analytic works cited, and presented here, considered purely
incompressible flow; for our simulations, we consider only very modest
compressibility, with projectile motions through the ambient fluid quite
subsonic, so it suffices for the moment to consider in the external
medium $\rho = \rho_0$.    The fluid here is further considered to be
infinitely conducting; however, the buildup of magnetic field without
a corresponding buildup of mass does not violate the `flux-freezing'
condition, as shown in the cartoon Fig.~\ref{fig:draping-cartoon} as
incoming fluid elements are `squished' along the sides of the incoming
sphere, so that the magnetic flux coming out the sides of the fluid
element remains constant, even as the concentration of field lines builds
up along the stagnation line.   Further increase in magnetic energy comes
from the stretching of field lines in the direction of motion of the core.

In reality, of course, the magnetic field does back-react onto the
flow, and the kinematic potential flow solution fails for two reasons
-- buildup of a strong magnetic field layer (which violates the
kinematic assumption) and creation of vorticity (in conflict with
the potential flow assumptions).

\begin{figure}[h!]
\centering
\plotone{\choosefigure{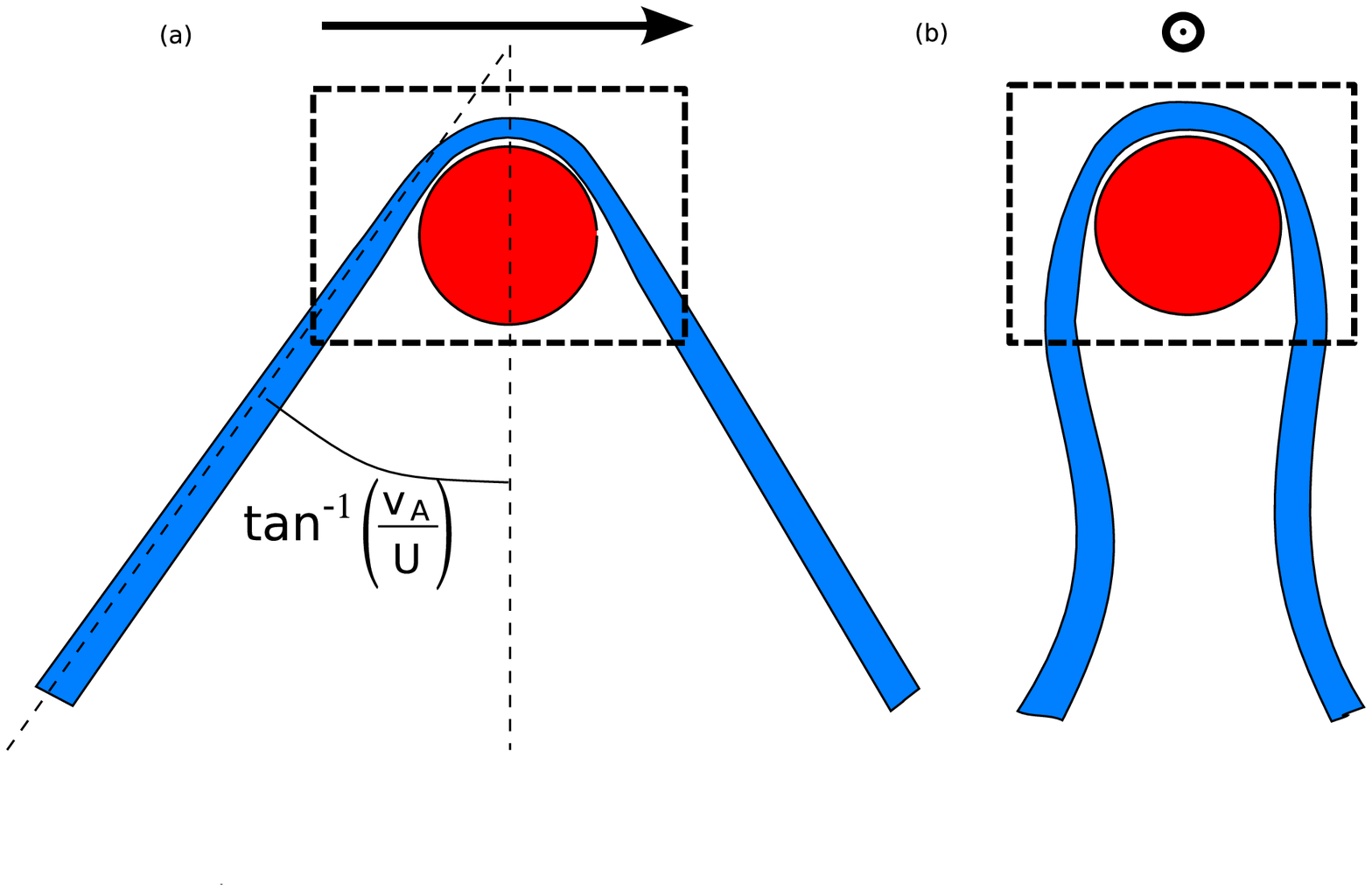} {drapegeometry.eps} {drapegeometry-bw.eps}}
\caption{A cartoon showing the expected geometry of the draped magnetic field
  (\ifallcolourfigs blue\else light\fi) over the object (\ifallcolourfigs
  red\else dark\fi).  Seen in the plane of the direction of the ambient field,
  panel (a), with the direction of the ambient field shown, a clean bow wave is
  presented with a well-defined opening angle. In the plane perpendicular to
  the ambient field, panel (b), the field lines can slip around the projectile,
  and the flow would close back in on a stagnation line on the other side of
  the object except for largely-2d vortical motions induced by instabilities at
  the magnetic interface.  The geometry of the flow in the region indicated by
  dashed box depends heavily on the final shape, and thus internal structure,
  of the moving object.}
\label{fig:drapinggeometry-cartoon}
\end{figure}

The magnetic field should exert a significant back-reaction when the
resulting magnetic pressure is comparable to the ram pressure of
the incoming material: $B^2/{8 \pi} \sim \rho_0 u^2$, where $u$ is
the speed of the core through the quiescent ambient fluid.   The
first place this will happen is along the stagnation line, which
by symmetry will be the location of the largest magnetic energy
density.   The layer of magnetic field with this magnitude is
expected (from Eq.~\ref{eq:stagline-bfield} and assuming $l \ll R$)
to be of thickness
\begin{equation}
l = \frac{1}{6 \alpha \MachAlfven^2} R
\label{eq:maglayer-thickness}
\end{equation}
where $\MachAlfven = u/\upsilon_A$ is the Alfv\'enic Mach number of the
core, $\upsilon_A^2 = B_0^2/{4 \pi \rho_0}$ is the ambient Alfv\'en speed,
and $\alpha$ is the constant of proportionality describing the
maximum magnetic pressure in units of the incoming ram pressure,
$B_{\mathrm{max}}^2/{8 \pi} = \alpha \rho_0 u^2$.   We will see
that $\alpha \approx 2$ and fiducial values for the situations
considered here will involve $\MachAlfven^2 \approx 3$, so that a
typical value for $l$ will be approximately $R/36$.   Even such
a very thin layer can have important effects, both in terms
of suppressing thermal conduction \citep{ettorifabian} and
hydrodynamic instabilities \citep{lintheory}.

It should be noted here that when we use the Alfv\'enic Mach number
$\MachAlfven$ through this work it should really be considered a
dimensionless ratio of ram pressure to magnetic pressure ($\MachAlfven^2 =
\rho_0 u^2/(2 P_{B_0})$), or at the least, some caution should be used
when interpreting it as a ratio of velocities ($u/\Alfvenvel$) as the
velocities are oriented in different directions; in the work presented
here, the velocity of the draped object will always be completely
orthogonal to the ambient direction of propagation of Alfv\'en waves.
Thus there is an important sense in which our projectiles are always
(infinitely) super-Alfv\'enic, which is not captured in the ratio
$\MachAlfven$.

Sweeping up such a magnetic field will occur on a timescale $t/t_c
\sim \sqrt{\alpha} (l/R) \MachAlfven \sim (\sqrt{\alpha} \MachAlfven)^{-1}$,
where $t_c = 2R/u$ is the projectile's own crossing time.   This
result means that, because the magnetic layer is very thin, a strong
field can be built up extremely quickly.   Crucially, particularly
for the propagation of bubbles, the sweep-up time can be significantly
smaller than a single crossing time; this is relevant because a purely hydrodynamic bubble
will generally self-disrupt into a torus, or smaller fragments in a
turbulent medium, in on order a crossing time \citep{flashbubbles,2007arXiv0709.1796P}.

\begin{figure}[h!]
\centering
\plotone{\choosefigure{pressure.comparison.log.panel.eps} {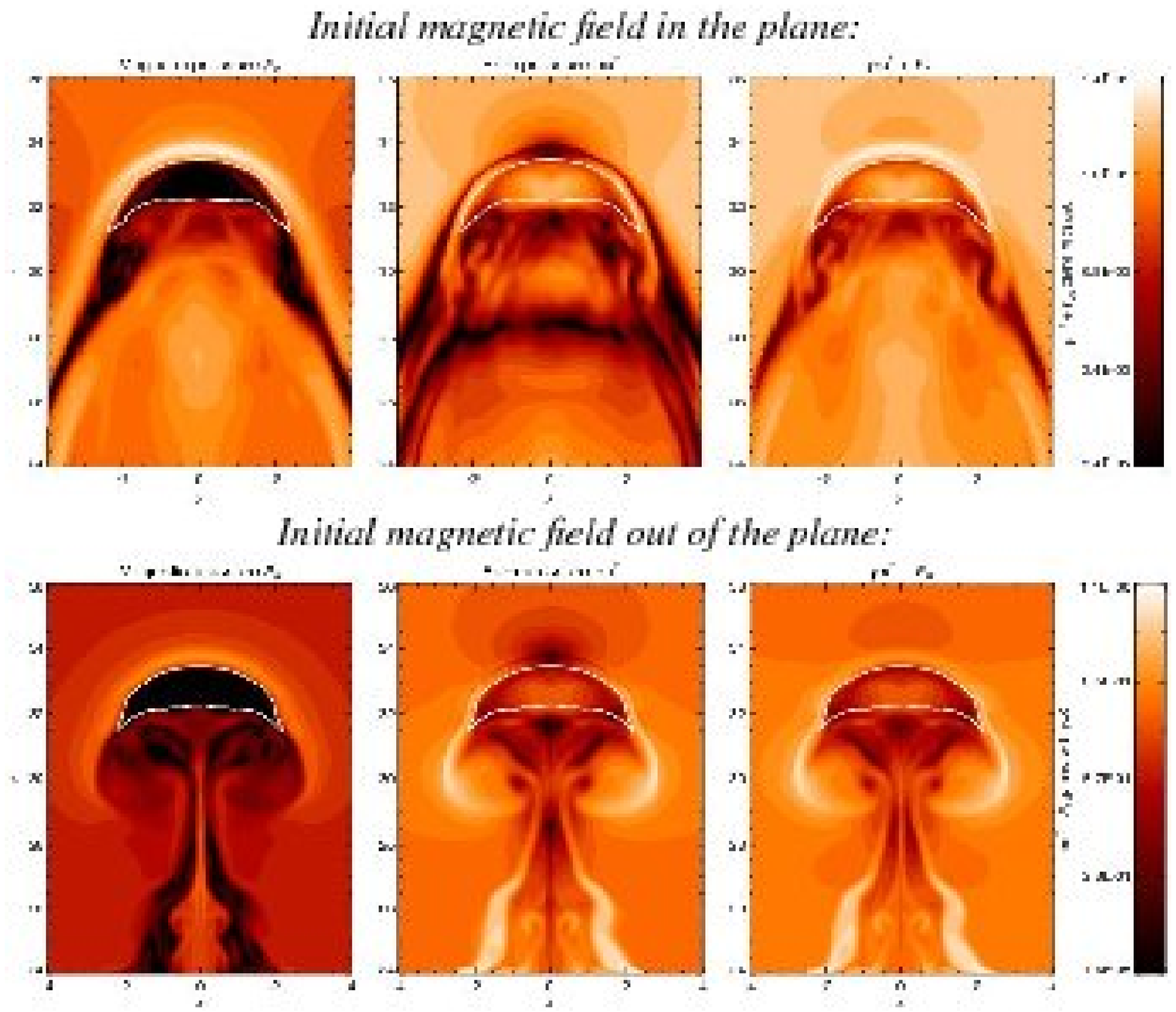} {pressure.comparison.log.bw.panel.eps}}
\caption{We compare the draped magnetic pressure and ram pressure in the plane
  that is parallel (perpendicular) to the initial magnetic field in the top
  (bottom) panels with a logarithmic color scale. In the parallel plane, the
  overpressure of the magnetic draping layer is only partly compensated by a
  deficit in ram pressure and eventually is responsible for decelerating the
  core due to magnetic tension. In the perpendicular plane, the ram pressure in
  the wake of the core attains much higher values and the draping layer closes
  towards the symmetry axis unlike the parallel plane where a nice opening cone
  forms. Shown here is a zoom-in on a small region of our computational domain.  \forhighresfiguressee}
\label{fig:mag/ram-pressure}
\end{figure}

This buildup of magnetic field will greatly effect the flow in the direction of
the ambient field lines, and the projectile will leave a magnetic bow wave
behind it; by analogy with other similar bow waves, we expect it to have an
opening angle of $\tan \theta \approx \upsilon_A / u$.  In the plane
perpendicular to the ambient magnetic field, however, the magnetic field will
have a much less direct effect as field lines can simply slip around the
projectile and instabilities can occur.  In the potential flow simulation, the
flow smoothly reattaches at the rear of the projectile; however, in this case,
vortical motions generated at the magnetic contact (where the magnetic pressure
and magnetic tension force is misaligned with the density gradient) and by
instabilities at the magnetic interface (which are not stabilized in this
plane) detach the wake from the object, leaving largely two-dimensional
vortical motions along the field lines in this plane.  The resulting expected
geometry is shown in a cartoon in Fig.~\ref{fig:drapinggeometry-cartoon} and
for our simulation in Fig.~\ref{fig:mag/ram-pressure}. This shows that the
draping layer becomes dynamically important and fills in the deficit of ram
pressure. The sum of the magnetic and ram pressure shows an over-pressure ahead
of the core that leads to a deceleration of the projectile.

\section{METHODOLOGY}
\label{sec:methodology}
In order to understand the full non-linear physics of magnetic draping around a
dynamically evolving dense projectile moving in a magnetized plasma we perform
our analysis in two steps. First, we analytically study the properties of the
flow of an ideally conducting plasma with a frozen-in magnetic field around a
sphere to explore the characteristics of the magnetic field near the surface of
the body.  To this end, we disregard any possible change in the flow pattern by
means of the back-reaction of the magnetic field. While the derivation of this
problem can be found in Appendix~\ref{sec:analytics}, we summarize the key
results in this section.  In the second step, we compare this analytical
solution to an MHD adaptive mesh refinement simulation and explore it
quantitatively.

\subsection{Analytical solution}
\label{sec:analytics_scetch}

The potential flow solution for an incompressible flow around a
spherical body reads as
\begin{equation}
\label{eq:v_solution_core}
\vel = \e_r \left(\frac{R^3}{r^3}-1\right) u\cos\theta +
     \e_\theta \left(\frac{R^3}{2 r^3}+1\right)u\sin\theta,
\end{equation}
where $R$ denotes the radius of the sphere and $u$ is the speed of the
core through the quiescent ambient fluid.  Using this solution, we
solve for the resulting frozen-in magnetic field while neglecting its
back-reaction onto the flow. For convenience, we show here the
approximate solution which is valid near the sphere,
\begin{eqnarray}
\label{eq:B_approx}
B_r &=& \frac{2}{3}B_0\sqrt{\frac{3s}{R}} 
         \frac{\sin\theta}{1+\cos\theta}\sin\phi, \\ 
B_\theta &=& B_0\sin\phi\,\sqrt{\frac{R}{3s}},\\
B_\phi   &=& B_0\cos\phi\,\sqrt{\frac{R}{3s}},
\end{eqnarray}
where we introduced a radial coordinate from the surface of the sphere, namely
$s=r-R$. These approximate solutions uniformly describe the field near the
sphere with respect to the angle $\theta$.  As described in
Appendix~\ref{sec:analytics}, the energy density of the magnetic field forming
in the wake behind the body is predicted to diverge logarithmically. We point
out that the validity of the potential flow solution heavily relies on the
smooth irrotational fluid solution where the magnetic back-reaction is
negligible. We will see that these assumptions are naturally violated in the
wake.

\subsection{Numerical solution}

\subsubsection{Setup}
\label{sec:numerics:setup}

\begin{figure}
\plotone{\choosefigure{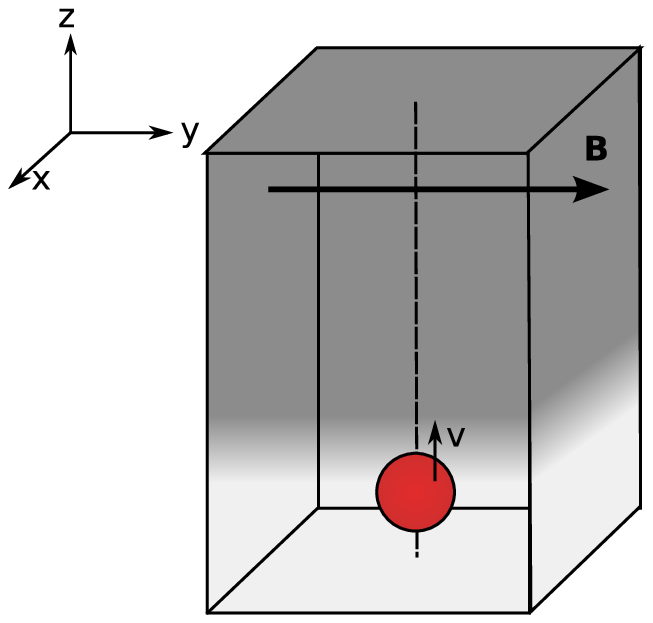} {simdomain.eps} {simdomain-bw.eps}}
\caption{Diagram showing the geometry of the simulations presented here.  A
  spherical projectile with a smooth density profile $\rho_{\mathrm{max}} (1 +
  \cos(\pi r/R))/2$ is sent in the $+z$ direction with an initial velocity $\v$
  through an ambient medium with density $\rho_0$ and a uniform magnetic field
  pointed in the $+y$ direction.  The magnetic field strength `turns on'
  through the domain with a $\tanh$-profile in the direction of motion of the
  projectile, as indicated by the shading of the box; this allows us to start
  the core in an essentially field-free region and smoothly enter the
  magnetized region.  Periodic boundary conditions are used in the directions
  perpendicular to the direction of motion, and zero-gradient `outflow'
  conditions are used in the $z$ direction.}
\label{fig:simdomain}
\end{figure}

The simulations presented in this paper are set up as shown in
Fig.~\ref{fig:simdomain}.   For clarity of understanding the physical
picture, we consider only the magnetohydrodynamics (MHD); no external-
or self-gravity is considered, and we defer other physics such as
self-consistent inclusion of thermal conductivity to future work.  In this
report, we also consider only the magnetic field of the external medium,
and assume that it is uniform over the scales of interest here.

In the code units we consider here, the ambient material has a density of
$\rho_0 = 1$, and a gas pressure $P = 1$.  The (unmagnetized) fiducial
projectile has a radius that we vary in our runs between $R = 0.5$ and 2, and a
maximum density of $\rho_{\mathrm{max}} = 750$.  With the density profile
chosen $\rho(r) = \rho_{\mathrm{max}}(1 + \cos(\pi r/R))/2$, the mean density
of the projectile is $(1 - 6/\pi^2) \rho_{\mathrm{max}}/2 \approx 0.2
\rho_{\mathrm{max}}$.  Both the ambient and projectile material are treated as
ideal, perfectly conducting fluids with ratio of specific heats $\gamma = 5/3$,
and so the adiabatic sound speed in the ambient medium is $\sqrt{5/3}$.  The
pressure inside the projectile is chosen so that the material is initially in
pressure equilibrium.

The projectile initial velocity is typically chosen to be $1/4$, for a
Mach number into the the ambient medium of $\approx 0.32$.  The simulation
in the transverse directions range from $[-4,4]$, and in the direction of
motion of the projectile ranges from $[0,28]$ for an aspect ratio of 2:7;
in most of the  simulations with projectiles larger than the fiducial $R =
1$, the domain size is increased proportionately.  The initial magnetic
field strength can be defined in terms of $\alpha_0 = \rho u^2/P_{B_0}$;
a typical value used in these simulations is $25/4$, or $P_{B_0} = 1/100$.
Periodic boundary conditions are used in the directions perpendicular to
the direction of motion, and zero-gradient `outflow' conditions are used
in the $z$ direction.  Experiments with different horizontal boundary
conditions produced no major differences in results.

The projectile fluid is initially tagged with a passive scalar, so that
the material corresponding to the projectile can be traced throughout
the simulation.

\subsubsection{Code Choice}
\label{sec:numerics:whyflash}

As can be seen from analytic arguments, and is shown in some detail
in \S\ref{sec:picture}, two features characterize the problem of
magnetic draping:  the formation of a narrow strongly-magnetized layer,
and the relative simplicity of the dynamics, in that a potential flow
solution with only magnetic field kinematics captures much of the problem,
lacking only the magnetic field back-reaction.

Because of the separation of scales (a relatively large object moving
through an ambient medium and a relatively small layer forming around
it), the highest resolution requirements would impose a large cost on
the simulations if the resolution had to be everywhere uniform; indeed,
it is only a small portion of the simulation domain which needs to be
resolved at the highest level.     This is especially true since the
simulations we will need to perform are three-dimensional (as we will see
in the next sub-section, it is impossible to do meaningful simulations
of magnetic draping in two dimensions).   Thus, a simulation code
which allows some adaptivity of meshing is extremely helpful for approaching
this problem.

The relatively straightforward magnetic field dynamics means that, unlike
in problems of (for instance) studying the details of MHD turbulence, we
do not require high-order finite difference methods; this is particularly
true because of the sharpness of the thin layers and the large density
gradients in this problem.   Instead, an MHD solver which can accurately
deal with sharp gradients is valued.

As a result of the importance of AMR for these simulations, 
the code we've chosen to perform these simulations with is the \FLASH{} code
\citep{flashcode,flashvalidation}.  \FLASH{} is an adaptive-mesh general
purpose astrophysical hydrodynamics code which is publicly
available\footnote{\url{http://flash.uchicago.edu}}.  The MHD solver we use here is a
dimensionally-split second-order accurate 8-wave Godunov-type solver which is described in more detail in
\cite{flashmhd}.  The smallness of spurious magnetic monopoles is ensured
by a diffusion-type `div-B' clean operation.  This diffusive cleaning
approach can be problematic near strong shocks, where diffusion
cannot operate quickly enough; however, no such shocks occur in these
simulations.  \FLASH{} has been often used for related problems such as
hot and magnetized bubbles in the intercluster medium
\citep[\eg{},][]{flashbubbles,2007arXiv0709.1796P,2006MNRAS.373L..65H,2006astro.ph.11531R,
  2006astro.ph.11444G,2005MNRAS.364...13P,2004MNRAS.355..995D,
  2003MNRAS.346...13H, 2003ApJ...592..839B,2002Natur.418..301B}.

\subsubsection{Parameters}

Performing simulations of draping over a projectile with an explicit
hydrodynamics code (so that compressibility effects will be included,
for ease of comparison with later, supersonic, work) with a finite
resolution places some restrictions on the range of parameter space
which can be explored.

For simulating these cases with no leading shock, we require that the velocity
of the projectile, $u$, be less than the sound speed in the ambient fluid --
but to take a reasonable number of timesteps (avoiding computational expense
and spurious diffusion) requires that the projectile velocity remain of the
order of the sound speed; thus $u \lesssim c_s$, or $\rho_0 u^2 \lesssim \gamma
P$, where $\rho_0$ and $P$ are the unperturbed density and pressure of the
ambient medium.  For the hydrodynamics of draping to be realistic, the magnetic
pressure in the fluid must be significantly less than the gas pressure,
$P_{B_0} \ll P$.  Finally, resolution requirements for resolving the thickness
of the magnetic layer will put some constraint on the thickness of the magnetic
layer $l > R/N$ from Eq.~\ref{eq:maglayer-thickness}, with $N$ being the number
of points which resolve the radius of the projectile; typically the size of the
domain (if at full resolution) will be $8N \times 8N \times 28N$.  This
constraint, expressed in terms of the relevant pressures (ram pressure and
initial magnetic pressure) $P_{B_0} \gtrsim 3 \alpha (\rho_0 u^2)/N$.
Combined, these constraints give
\begin{equation} \gamma P \gtrsim \rho_0 u^2 \gg P_{B_0} \gtrsim
\frac{3 \alpha \rho_0 u^2}{N} \label{eq:simulationconstraints}
\end{equation} For a given $u$ -- which is more or less arbitrary,
fixed to be near the (arbitrary) sound speed -- there is thus a
relatively narrow range of initial magnetic pressures in terms of
the ram pressure of the ambient material onto the projectile which
can be efficiently simulated.   As we will see, $\alpha \approx 2$,
and for the simulations presented here, $N \sim 32 - 64$, meaning
we are constrained to study roughly that part of parameter space
where $\rho_0 u^2 / P_{B_0} \approx 1 - 10$.

\subsubsection{Comparison to Previous Numerical Work}

Having described the setup of the simulations we perform for this
work, it is worth comparing this to previous numerical work.  One body
of work \citeeg{miniati1,miniati2,miniati3} considered a very similar
problem in the context of the ISM, where a dense cloud may be moving
through (for instance) the magnetic field associated with the Galactic
spiral arms.  In this context, a cloud of cold-phase ISM gas may very
well not be globally self-gravitating (or only weakly so) and so a
roughly `top-hat' density function is used to describe the cloud.  In
addition, as appropriate to the ISM, supersonic motion was imposed on
the ambient material, with the cloud experiencing a high-Mach number
shock and an ambient magnetic field at the same time; this might be
appropriate for instance as a cold-phase cloud encounters a hot-phase
medium for the first time.  In our work, we are considering subsonic
motion initially, so that one significant difference is the flow speed
of the draping material, in addition to the much flatter density
profile used in the previous work.  While we defer the supersonic case
to future work, we note that in our context, for example in the case
of (for instance) a core falling into a cluster, we would expect the
core to slowly accelerate before encountering significant magnetic
field, so the double transient of being shocked and exposed to
magnetic field at the same time would not be appropriate for the case
we wish to consider.

Other more recent work, which in large part motivated the research
presented here, considers squarely the case we are interested in --
magnetic draping in cluster environments \cite{asai04,asai05,asai07}.
In these works, the authors similarly consider a spherical projectile
moving through an ambient magnetized medium.  These simulations
include more physics (\eg{} anisotropic thermal
conduction, tangled magnetic fields) than we consider here, as the
goal of these works was to find out if the draping of the magnetic
field could significantly suppress thermal conduction into a cold core.
These simulations show that magnetic draping is potentially important
for suppressing thermal conduction across the interface, despite the
fact that their draping layers are under-resolved by at least a factor
of 100 (cf.{} Eqns.~\ref{eq:maglayer-thickness} and
\ref{eq:simulationconstraints}).

In a slightly different cluster environment, MHD effects can also
stabilize the morphologies of buoyantly rising radio bubbles from AGN
in cluster centers. \citet{ruszkowskiI} carefully set up magnetically
isolated bubbles and tangled cluster magnetic fields in the ambient
medium as required for causally or temporally disconnected generating
processes of these fields. They showed that magnetic draping of
cluster magnetic fields possibly supported by helicity of magnetic
fields internal to the bubble can strongly alter the dynamics and
suppress hydrodynamic instabilities; thus resembling the observed
morphologically intact X-ray cavities.

Knowing that under some circumstances thermal conduction can be
suppressed by magnetic draping and hydrodynamic stability can be
enhanced, it becomes worthwhile to understand the process of draping
itself in some detail, in which case the additional micro-physics (at
this stage) makes understanding this basic process more, not less,
difficult.  Thus we consider here a case with no explicit
micro-physical conduction and simplified magnetic field geometry. With
this choice, our simulations remain scale free and can be applied to
various astrophysical environments.  Additionally, our simulations can
be directly compared to analytics and we are able to carefully examine
the process of the draping itself.  Further, the more simplified
physics along with the use of AMR allows us to run simulations for
significantly longer times as the more complicated earlier simulations.

\subsubsection{Two Dimensional Results}
\label{sec:numerics:boing}

In two dimensions, the imposition of a symmetry greatly limits
possible magnetic field geometries.  In an axisymmetric geometry,
the only meaningful uniform field geometry is parallel to the axis of
symmetry, which in this case would also be the direction of motion
of the projectile; in this somewhat artificial case magnetic field
could somewhat constrain a projectile (or a bubble; \cite{flashbubbles})
but draping could not occur.

In planar symmetry, the field can have components out of the plane,
in the plane parallel to the direction of motion, or in the plane
perpendicular to the direction of motion.    A component out of the
plane will only have the dynamical effect of adding to an effective gas
pressure (\citeeg{chandra}); the component along the direction of motion
of the projectile cannot be draped.

\begin{figure}
\centering
\plotone{\choosefigurea{bounce-labelled.eps} {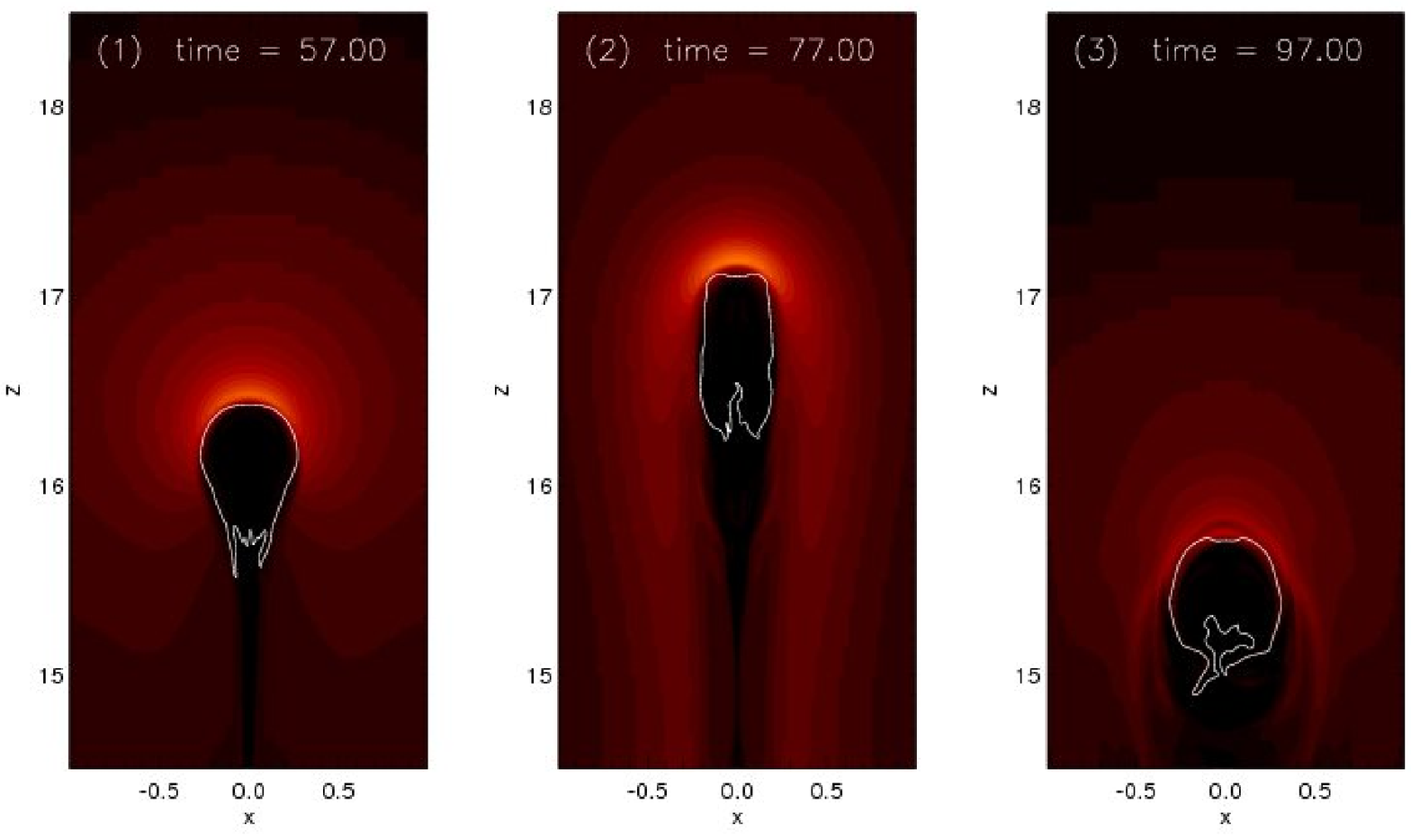} {bounce-labelled-bw.eps}}
\epsscale{.5}
\plotone{\choosefigureb{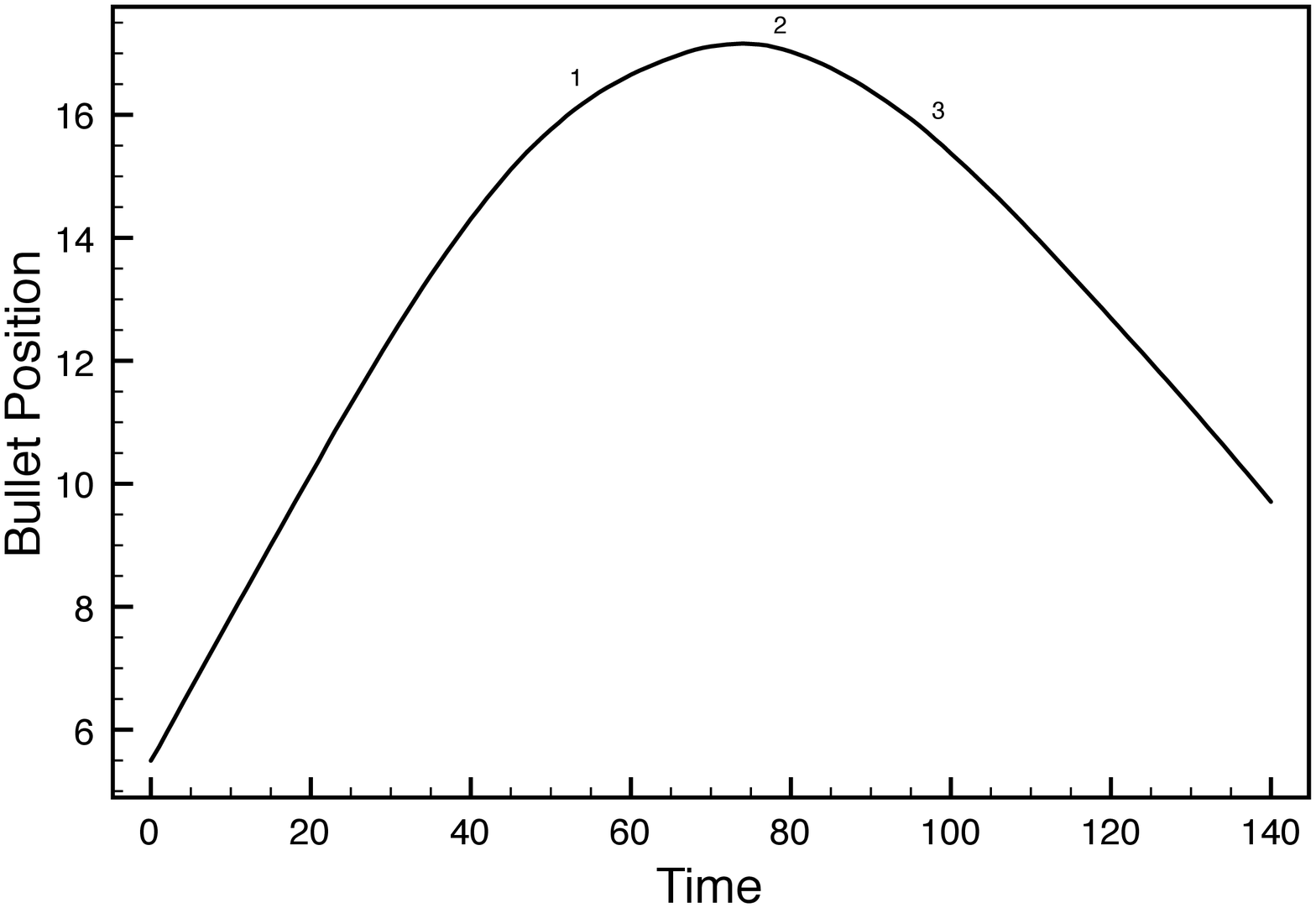} {bounce-2d-lineplot.eps} {bounce-2d-lineplot.eps}}
\epsscale{1}
\caption{The above is a plot of the projectile position (calculated
  here by the maximum height at which there is significant projectile
  material at any given time) over time in a two dimensional draping
  simulation.  At about time 75, the projectile is
  actually bounced back under the extremely strong magnetic tension
  which in two dimensions must grow ahead of the projectile.
  The top panel shows $B^2$ in a closeup of the simulation domain at three
  representative times during the bounce, with a white contour indicating
  the position of the original projectile material.  \forhighresfiguressee}
\label{fig:2dresults-plot}
\end{figure}

Previous work \citeeg{asai05} has examined the case with two dimensional
planar symmetry with a magnetic field in the plane of the simulation and
perpendicular to the direction of motion of the projectile.   However,
in this case, field lines cannot cannot slip around the projectile, and so as
more and more magnetic field gets swept up by the projectile, magnetic
tension grows monotonically and linearly ahead of the projectile until the
forces becomes comparable not only to the ram pressure seen by the projectile
of the ambient medium, but to the ram pressure of the projectile as seen by
the ambient medium.  At this point, the projectile trajectory is reversed.
A figure describing this is shown in Fig.~\ref{fig:2dresults-plot}, where
the magnetic tension forces are seen to compress the projectile (with mean
density $\approx 150$ times that of the magnetized medium) before repelling
it.

This outcome is hinted at in Fig. 4 of \cite{asai05}, where in 2d the
magnetic energy increases linearly and without bound, while the 3d models
reach a maximum magnetic energy.

\subsubsection{Three Dimensional Simulations}
\label{sec:numerics:3dsims}

A listing of the eight main runs done for this work are shown in
Table~\ref{tab:runs}, and the basic setup follows the discussion earlier.
The parameters varied are the size of the projectile, its velocity, and
the strength of the ambient magnetic field.  Other runs (the equivalent
of run B but with half the resolution, or with the same resolution
but differing boundary conditions) were run to confirm that the results
did not change; they are not listed here.

\begin{table}[ht!]
\centering
\begin{tabular}{cccccc}
\hline
\hline
Run & $R$ & $u$ & $P_{b,0}$ & $\rho u^2/P_{b,0}$ & $R/\Delta x$ \\
\hline
A & $\frac{1}{2}$ & $\frac{1}{4}$ & $\frac{1}{100}$ & $6.25$  & $64$ \\   
B & $1$           & $\frac{1}{4}$ & $\frac{1}{100}$ & $6.25$  & $64$ \\   
C & $2$           & $\frac{1}{4}$ & $\frac{1}{100}$ & $6.25$  & $32$ \\   
D & $1$           & $\frac{1}{4}$ & $\frac{1}{50} $ & $3.125$ & $64$ \\    
E & $\frac{1}{2}$ & $\frac{1}{8}$ & $\frac{1}{100}$ & $1.5625$& $32$ \\   
F & $\frac{1}{2}$ & $\frac{1}{4}$ & $\frac{1}{100}$ & $6.25$  & $32$ \\   
G & $\frac{1}{2}$ & $\frac{1}{2}$ & $\frac{1}{100}$ & $25$    & $32$ \\   
H & $2$           & $\frac{1}{4}$ & $\frac{1}{250}$ & $15.625$& $128$ \\   
\hline
\end{tabular}
\caption{Details of 3-dimensional simulations run for this work.  Simulations
were run with an ambient density and pressure of $1$ in code units, and $\gamma = 5/3$. 
Simulations were run until maximum magnetic field on stagnation line was approximately
constant, typically 40-80 time units.}
\label{tab:runs}
\end{table}

\begin{figure}
\centering
\plotone{\choosefigure{stagnation-convergence-b.eps} {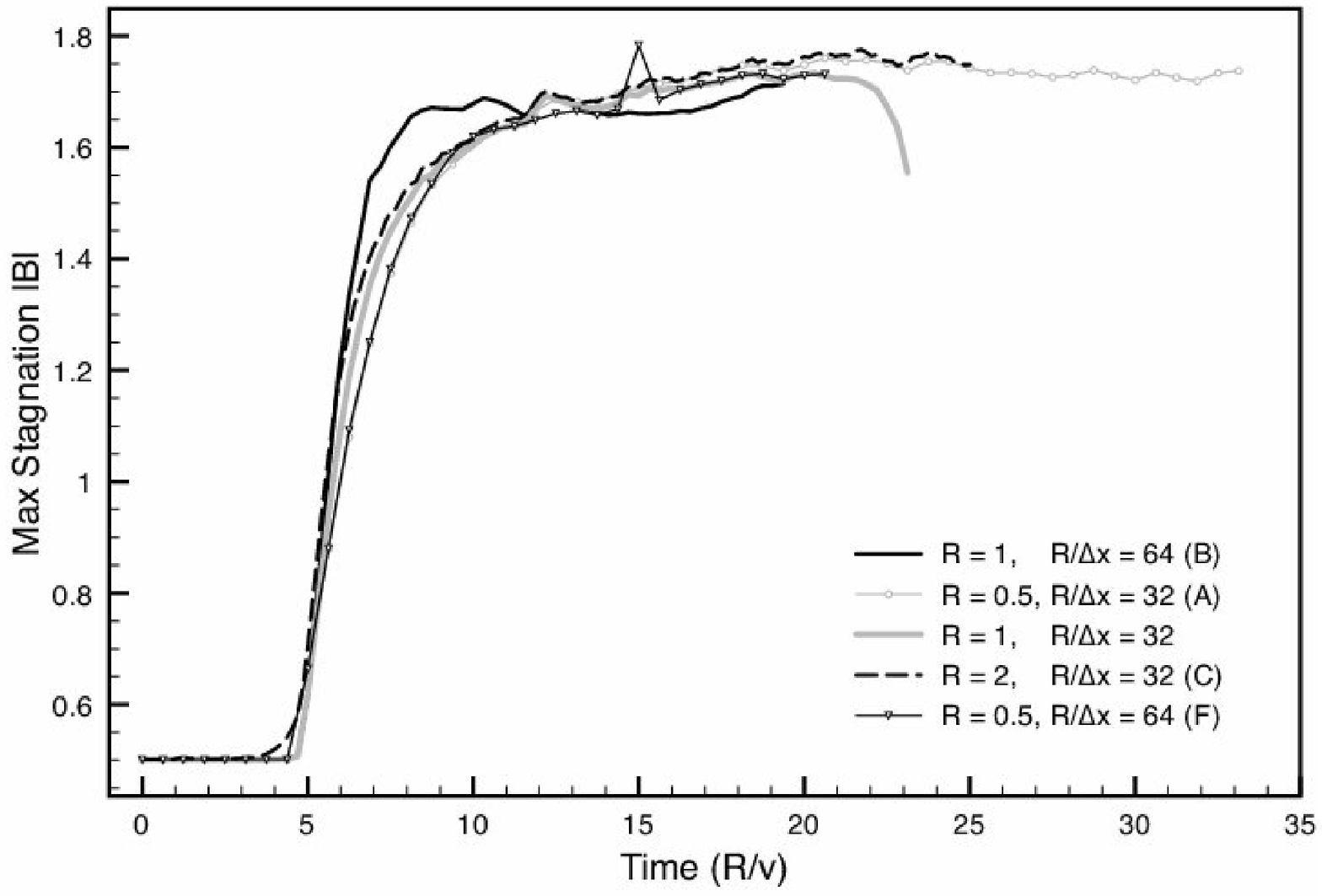} {stagnation-convergence-b.eps}}
\caption{A plot of the maximum magnetic field strength along the
stagnation line for runs with three different projectile sizes and two
different resolutions across the projectile, all with the same velocity into
the ambient medium and the same ambient magnetic field.    Plotted is
the maximum magnitude of the magnetic field along the stagnation line
vs the time (scaled to the crossing time of the projectile).   The projectile
initially sits in an unmagnetized region.   The maximum magnetic field
strength along the stagnation line is a sensitive measure of whether
the structure of the draped magnetic field is being resolved; we see
here clear evidence that with the resolution used in this simulation
the draped layer is being adequately resolved.   In the low-resolution $R=1$
simulation, which also was run in a somewhat smaller box, towards the end of the run the draping layer begins to 
leave the top of the simulation domain, leading to the sudden rapid drop
in magnetic field strength.
}
\label{fig:stagnation-convergence}
\end{figure}

To show that these runs were producing results independent of resolution,
the maximum magnetic field strength along the stagnation line for all the
runs with the same ${P_B}_0$ and $\rho u_0^2$ are plotted versus time
in Fig.~\ref{fig:stagnation-convergence}.   The maximum field strength
is sensitive to the resolution of the magnetic field layer, but we see
here that varying the resolution by a factor of two does not effect the
results, as the field layer is adequately resolved (but only marginally
in the case of $R/\Delta x = 32$).    We also see, as we'd expect from
the discussions in \S\ref{sec:picture}, that the field strength in the
layer does not depend on the size of the core.

The magnetic layer in Run G is under-resolved; while the value of
$R/\Delta x$ is the same as other runs, the velocity is higher, so that
by Eq.~\ref{eq:maglayer-thickness} the layer is thinner.  We include
this run because it demonstrates certain robustness of results; although
the layer structure is not adequately resolved at the stagnation point,
other global properties of the magnetic layer (geometry and dynamical
effects) otherwise remain robust.

Indeed, one should be careful about what one means by `resolved'.
This discussion should not be taken to mean that the other
simulations are in all respects resolved.   In particular, as we
will see in \S\ref{sec:comparison} and \S\ref{sec:characteristics},
and as suggested by \cite{lintheory},  the flow around the bubble
in the $xz$ plane (\eg{}, transverse to the initial magnetic field)
is unstable to Kelvin-Helmholtz and Rayleigh-Taylor instabilities,
not stabilized by the presence of magnetic field.    Since we have not
prescribed any small-scale dissipative physics in these simulations,
these instabilities will never properly converge with increased resolution
\citeeg{flashvalidation} as new unstable scales are added.  This can only
be corrected by adding small-scale physics, \eg{} thermal diffusion; this
is left for future work, as the relevant microphysics is itself a current
research problem \citep{lyutikovdissipation,schekochihindissipation}.
While the rate of development of the instabilities and their properties
is very important for long-term mixing of the material of the moving
object with that of the surrounding medium, we restrict ourselves here to
studying the development of the magnetic layer and its global properties.

A note here on our AMR is in order.  In Table~\ref{tab:runs}, we show the
finest resolution achieved in the simulation.   \FLASH{} places resolution
elements based on a second-derivative refinement criterion, in an attempt
(appropriate for a second-order accurate code) to reduce error; see
\cite{flashcode} for more details.   Here we refined based on pressure,
density, and composition (\eg, fraction of projectile material); early
experiments with also refining on magnetic field components showed little
or no difference.   The result is that the entire region immediately
surrounding the projectile is fully refined, and the wake region is
nearly fully refined.   However, because no interesting flow features
are occurring in most of the domain at any given time, the savings from
using AMR can be substantial; for run G, for instance, the very simple,
laminar initial conditions require only $9.7 \times 10^5$ zones; by
time 40, this has increased to $\approx 8.1 \times 10^6$ zones, and by
time 75, an approximately-steady $9.1 \times 10^6$ zones are being used;
this is to be compared to the $469.8 \times 10^6$ zones, or a factor of 51 greater,
that would be required to resolve the entire domain at this same finest
resolution.

A final feature worth noting in our runs is that for run E, $\MachAlfven^2 =
1/2 (\rho u^2 / {P_B}_0) \approx 0.78 < 1$;  that is, this run has the
projectile moving sub-Alfv\'enically, if only marginally.   However,
because of the field geometry, we will see that this makes essentially no
difference for the  draping.   This is simply because the Alfv\'en speed
in the direction of motion of the projectile is zero -- no component of
the magnetic field points in that direction.   While the exact imposition 
of this condition in our initial conditions is somewhat artificial in
this case, it is always true that it is only the component of the magnetic
field that lies transverse to the direction of motion that will be draped.

\section{COMPARISON OF THEORY AND SIMULATIONS}
\label{sec:comparison}
\subsection{Magnetic field along the stagnation line}

The first comparison we make is to the one-dimensional predictions made
along the stagnation line, for instance in \cite{lyutikovdraping},
where a very specific prediction is made for the ramp up, with a
particular functional form, of the magnetic field strength given in
Eq.~\ref{eq:stagline-bfield},   and it is not necessarily clear that
such a prediction will hold when the projectile begins to deviate
significantly from spherical.

\begin{figure}[h!]
\centering
\plotone{\choosefigure{lessrefine-big-30.eps} {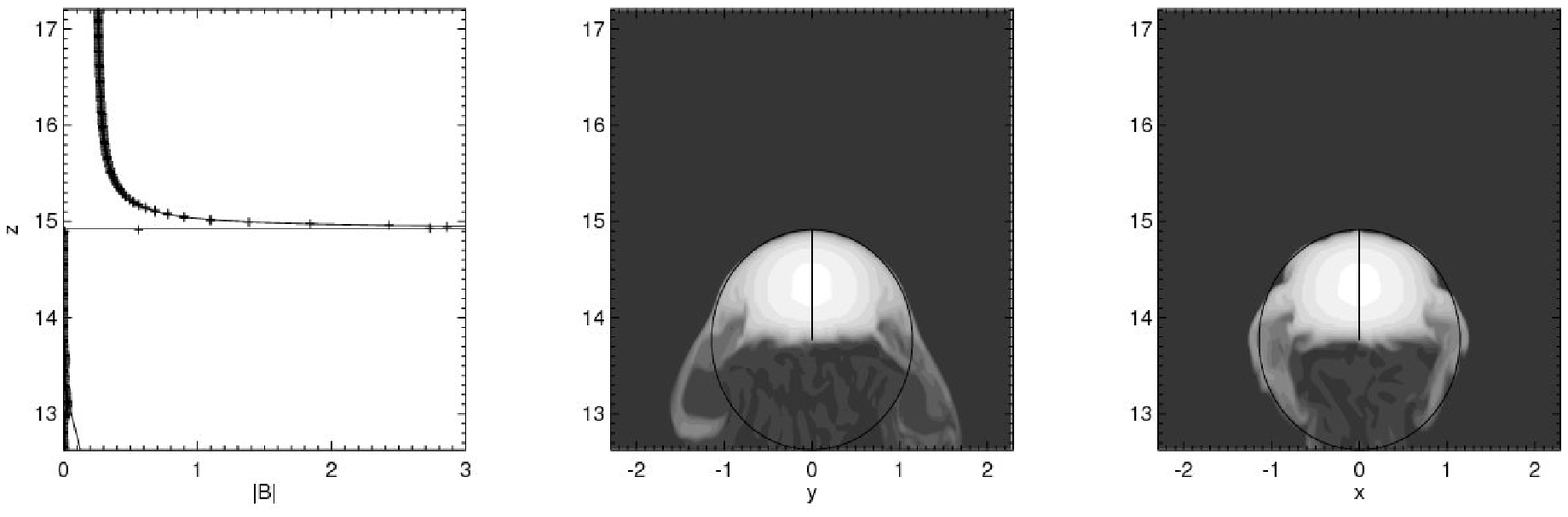} {lessrefine-big-30.eps}}
\caption{Shown is, left, the magnetic field along the stagnation line
in the simulation ('+') and a fitted theory prediction, with the two
fitting parameters being the position of the peak and a radius giving
a characteristic fall off of the field strength.   On the right are
cut-planes along and across the initial magnetic field of the density of
the projectile, with a circle of radius and position given by the fit to the
magnetic field structure, left.   The radius given by the fit corresponds
with the radius of curvature at the working surface of the projectile.
Results are taken from run B at time $t = 38.75$; results from other
simulations and other times give similarly good fits.  \forhighresfiguressee}
\label{fig:sidebyside}
\end{figure}

To make this comparison, we extract the magnetic field strength along
the stagnation line for an output, and fit it to the equation
\begin{equation}
\frac{|B|}{\rho} = \frac{B_0}{\rho_0} \frac{1}{\sqrt{1 - \left ( \frac{R}{(z-z_0)} \right )^3 }}
\label{eq:fitstagline}
\end{equation}
where $B_0$, $\rho_0$ are known, and the fit is for the parameters $R$,
which would correspond to the radius of the sphere, and $z_0$, which
would be the position of the centre of the sphere.   Results for a typical
output are shown in Fig.~\ref{fig:sidebyside}.   Not only do the fits well
represent the behaviour magnetic field strength, but they also suggest
a physical interpretation for the functional forms interpretation even
when the projectile becomes significantly non-spherical; $R$ becomes
the radius of curvature of the working surface of the projectile at the
stagnation line.   As the projectile becomes more distorted, $R$ can
become significantly larger than the initial radius of the projectile;
in this example, the expansion is a relatively modest 15\%.

\subsection{Comparison of the velocity field}

\begin{figure}
\centering
\plotone{\choosefigure{comparepotential.eps} {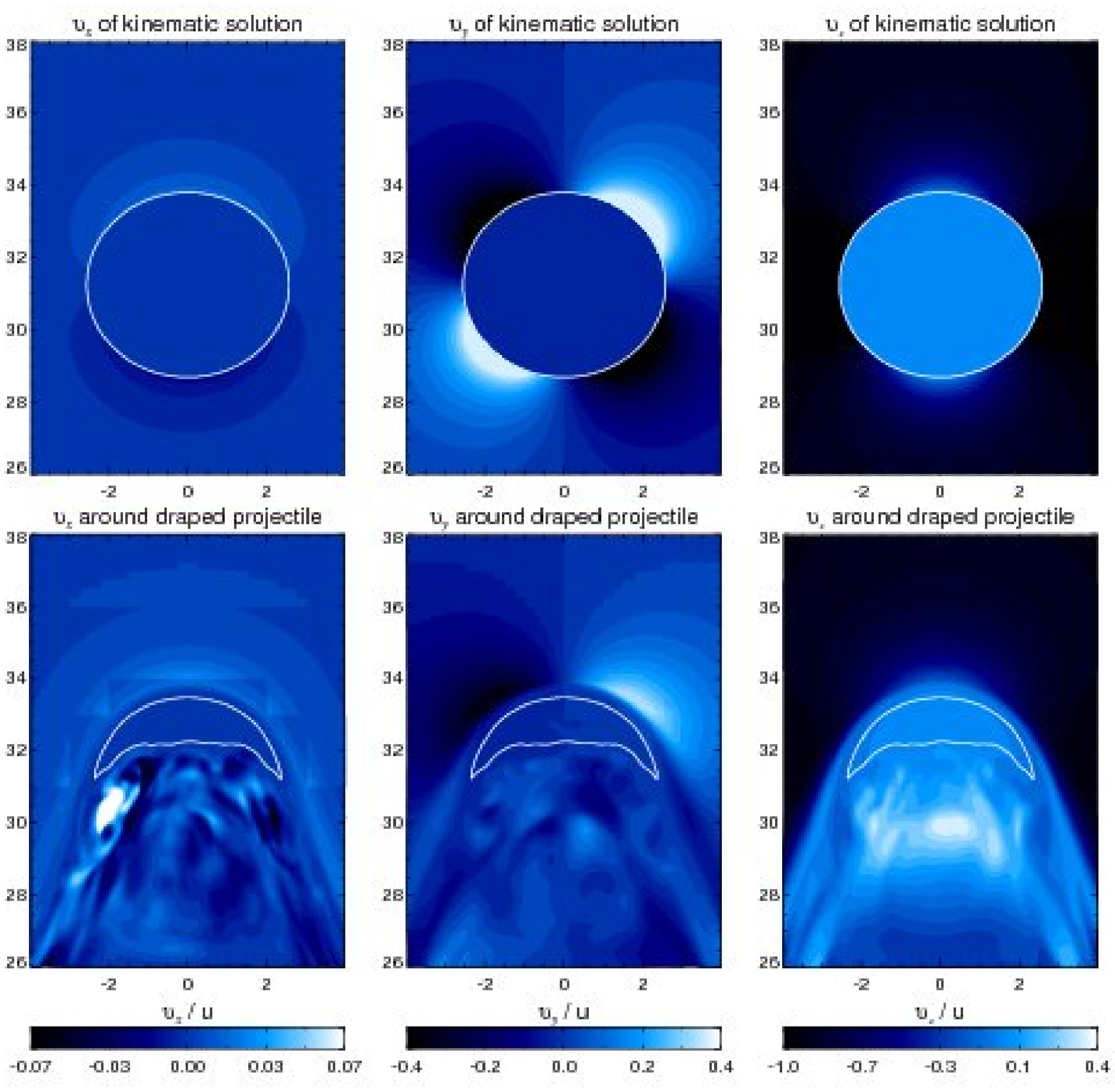} {comparepotential.bw.eps}}
\caption{Comparison of the velocity field of the analytical solution in the
  kinematic approximation (top panels) with our numerical simulation (bottom
  panels) in the plane of the initial magnetic field. While the velocity fields
  resemble each other very well in the upper half-space, there are distinct
  differences in the lower half-space.  These are due to the non-linear
  back-reaction of the dynamically important magnetic field in the draping
  layer on the MHD flow that generates vorticity in the wake of the projectile
  (cf.{\ }\S\ref{sec:vorticity}). Shown here and in the next figures is a
  zoom-in on a small region of our computational domain that extends up to 112
  length units and is four times larger in $x$ and $y$ direction. Note that we
  symmetrized the color map of the $\v_x$-component in order not to be
  dominated by one slightly larger eddy.  \forhighresfiguressee}
\label{fig:compare_vel}
\end{figure}

\begin{figure}
\centering
\plotone{\choosefigure{compareRhoVsquared.eps} {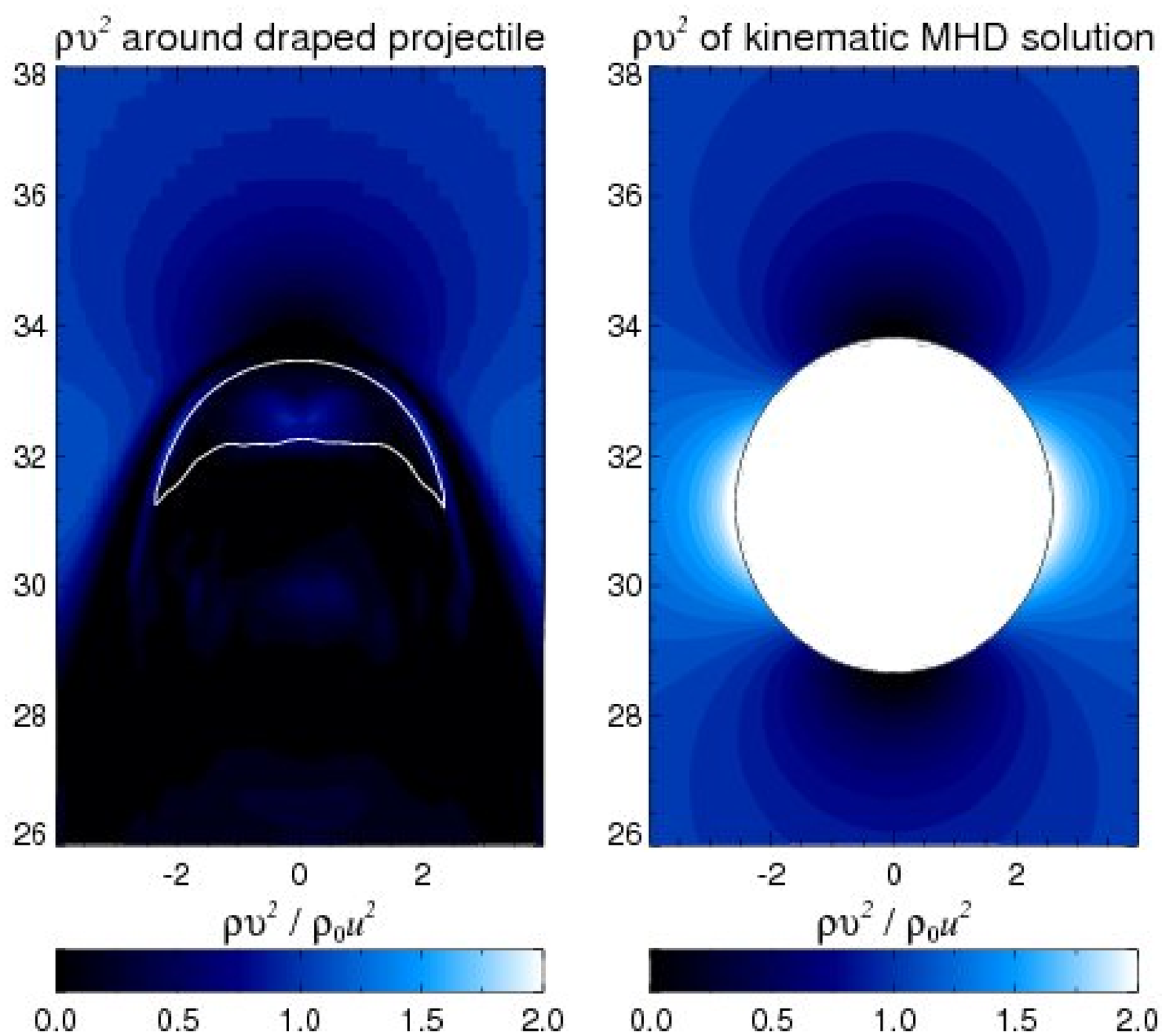} {compareRhoVsquared.bw.eps}}
 \caption{Comparison of the ram pressure in our numerical simulation
  (left panel) with the analytical solution in the kinematic
  approximation (right panel). Ahead of the projectile, the ram pressure
  resembles an exact potential flow behavior up to the draping layer
  which can be seen as a black layer around the projectile with a deficit
  of hydrodynamical pressure. Non-linear magnetic back-reaction of the
  draping layer causes the flow to depart from the potential flow
  solution and to develop vorticity.  \forhighresfiguressee}
\label{fig:compare_RhoVsquared}
\end{figure}

To compare the potential flow calculations, done in the frame of the
spherical body, with those of the numerical simulations, we transform the
numerical simulations into the frame of the projectile.  Because the
projectile slows down over time, we do not use the initial velocity
$u_0$ for this transformation, but measure the instantaneous
mass-weighted velocity of the projectile, by making use of the fact
that we are tracking the fluid that initially resided in the projectile
by use of an advected passive scalar, $a$.  Thus we measure the
instantaneous velocity of the projectile as
\begin{equation}
u = \frac{\bra\rho a \v_z\ket}{\bra\rho a\ket}.
\label{eq:bulletvelocity}
\end{equation}

The top panels of Fig.~\ref{fig:compare_vel} show the analytical solution of
the velocity field around the spherical body with radius $R$ in the kinematic
approximation.  For convenience and to simplify the comparison to the magnetic
field visualization, we show the Cartesian components of the velocity field. At
infinity, the fluid is characterized by a uniform velocity $\vel=-\e_z\, u$.
The quadrupolar flow structure results from the fluid decelerating towards the
stagnation line, the successive acceleration around the sphere until
$\theta=\pi/2$ and mirroring this behavior in the lower half-plane by symmetry.
The bottom panels of Fig.~\ref{fig:compare_vel} show the numerical solution of
the velocity field around an initially spherical projectile that deformed in
response to the non-linear evolution of the magnetized plasma. The white line
reflects the 0.9 contour of the `projectile fluid' and corresponds to an
iso-density contour.\footnote{Note that the apparent grid structure seen in the
  upper part of our simulated $\v_x$-component is an artifact of our plotting
  routine as well as small grid noise.  The interpolation scheme of the
  plotting routine falsely interpolates a smooth velocity gradient with an
  entire AMR block while it actually drops quickly to the velocity value at
  infinity.}  The quadrupolar flow structure in the upper half-plane resembles
nicely the analytic potential flow solution. As the flow approaches the
projectile and surrounds it, there are important differences visible. In the
analytical solution, the flow accelerates for $\theta \leq \pi/2$ and
decelerates for larger angles $\theta$. In the numerical solution, the magnetic
draping layer is stationary with respect to the projectile. This causes the
flow almost comes to rest in the magnetic draping layer. The back-reaction of
the magnetic draping layer on the flow casts a `shadow' on the wake of the
projectile. It prevents the flow to converge towards the symmetry axis and
suppresses the deceleration of the flow. Instead, vorticity is generated at the
draping layer which will be studied in detail in \S\ref{sec:vorticity}. The
comparison of the ram pressure in Fig.~\ref{fig:compare_RhoVsquared} underpins
this argument.

\subsection{Comparison of the magnetic field}

\begin{figure}
\centering
\plotone{\choosefigure{comparemagnetic.panel.eps} {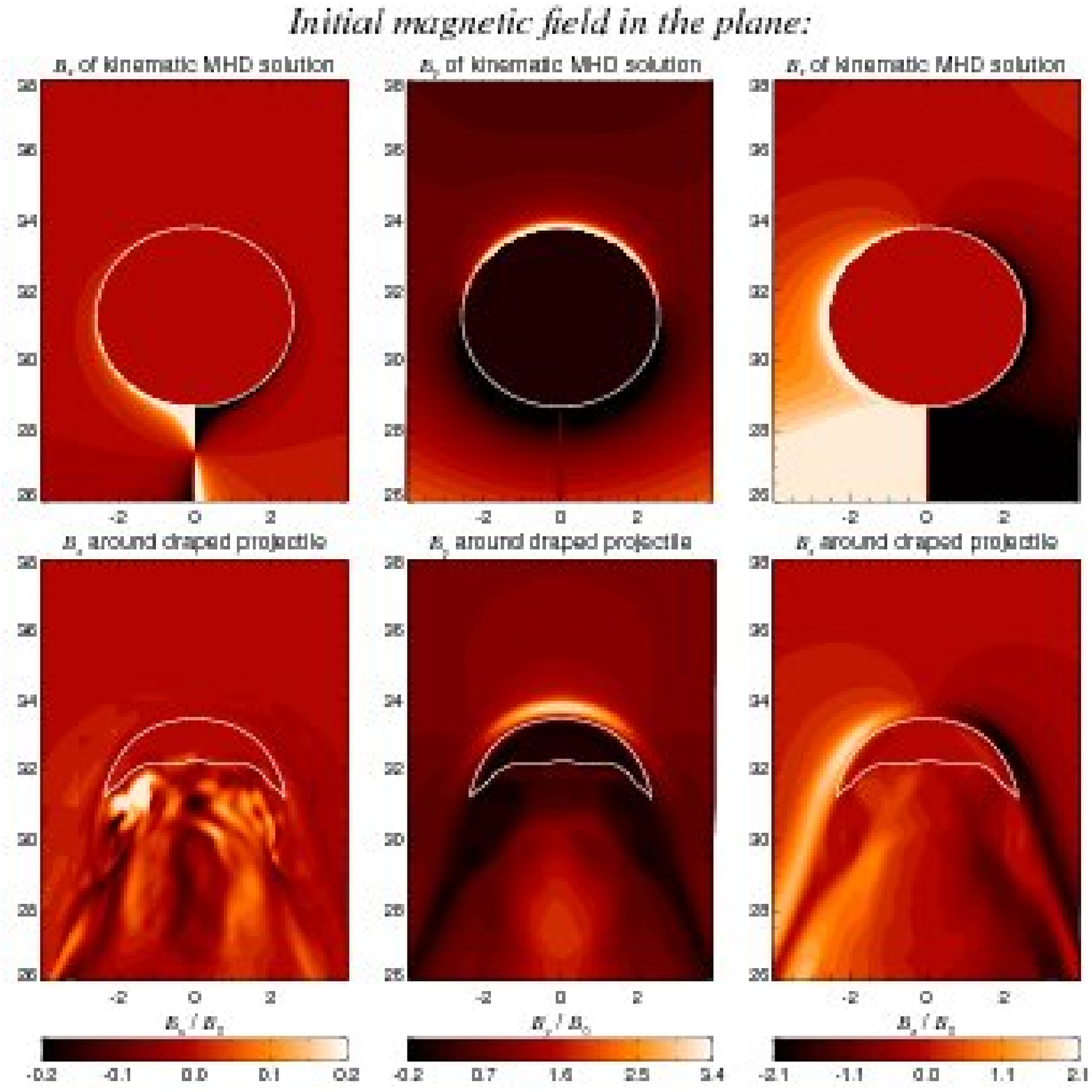} {comparemagnetic.bw.panel.eps}}
\caption{Comparison of the magnetic field in our numerical simulation (bottom
  panels) with the Taylor expansion of the analytical solution in the kinematic
  approximation that strictly applies only near the sphere (top panels). We
  show the Cartesian components (left to right, $x, y, z$) of the magnetic field in the plane that is
  parallel to the initial magnetic field.  There is a nice agreement between
  both solutions in the upper half-space, while there are again distinct
  differences in the lower half-space. The magnetic shoulders behind the projectile
  can be identified that prevents the draping layer from contracting towards
  the symmetry axis. In addition, MHD turbulence starts to develop in the wake
  of the projectile. \forhighresfiguressee}
\label{fig:compare_mag}
\end{figure}

\begin{figure}
\centering
\plotone{\choosefigure{comparemagnetic.perpendicular.panel.eps} {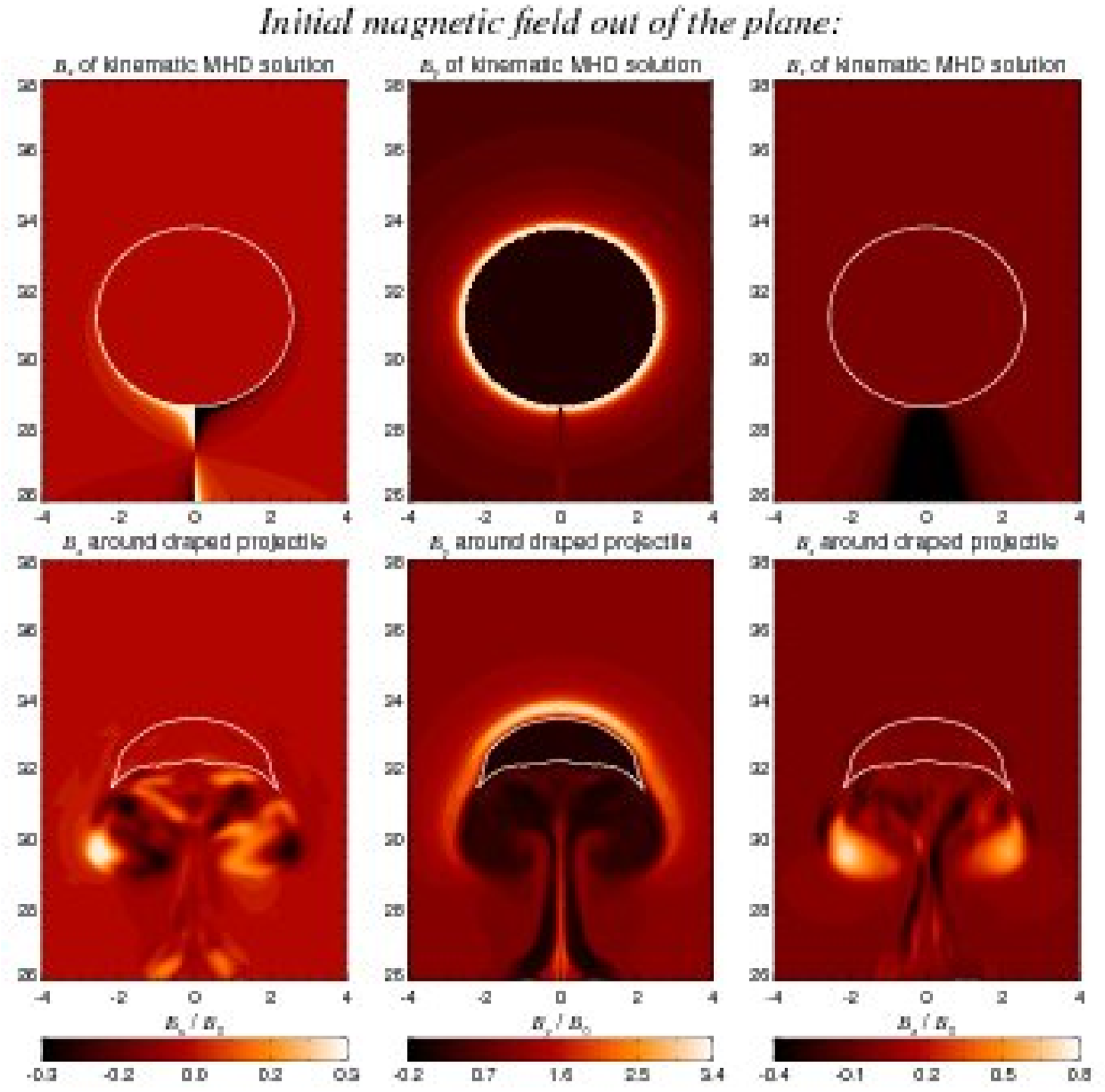} {comparemagnetic.perpendicular.bw.panel.eps}}
\caption{Same as previous figure, but in the the plane that is perpendicular to
  the initial magnetic field.  Shown is, left to right, the $x, y, z$ components. As expected from our analytic solutions, the
  draping layer forms by piling up magnetic field lines ahead of the projectile.
  The irregular magnetic field in the wake is generated by the vorticity that
  is absent by definition in our potential flow solution.  \forhighresfiguressee}
\label{fig:compare_mag_perp}
\end{figure}

\begin{figure}
  \begin{center}
  \epsscale{.5}
    \plotone{\choosefigure{compareBsquared.panel.eps} {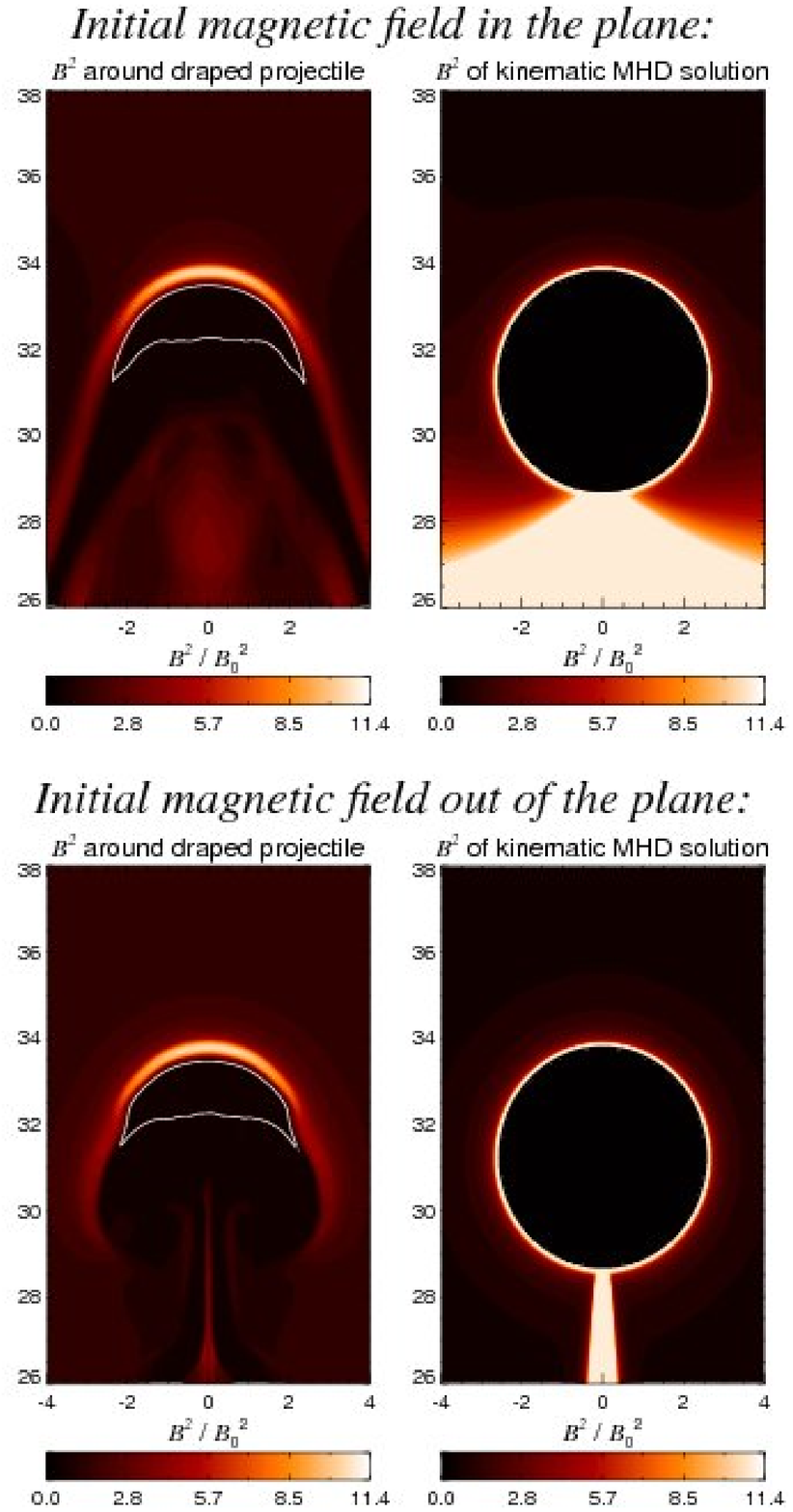} {compareBsquared.bw.panel.eps}}
  \epsscale{1}
  \end{center}
\caption{Comparison of the magnetic energy density in our numerical simulation
  (left panel) with the analytical solution in the kinematic approximation
  (right panel). The top (bottom) panels show the plane that is parallel
  (perpendicular) to the initial magnetic field.  In the analytical solution
  there is a narrow magnetic layer draped around the spherical body, while in
  our simulations the draping layer peels off behind the projectile due to
  vorticity generation. The geometry of the magnetic draping layer in the upper
  half-plane is very similar in both planes suggesting there an approximately
  spherical symmetry.  In the wake of the projectile, the draping layer forms a
  characteristic opening angle while the field lines can swipe around the
  projectile in the perpendicular plane and the draping layer closes towards
  the symmetry axis. \forhighresfiguressee}
\label{fig:compare_Bsquared}
\end{figure}

It is instructive to compare the analytic solution of the frozen-in magnetic
field in the kinematic approximation to the numerical solution in the planes
that are parallel and perpendicular to the initial magnetic field. We compare
the individual Cartesian components of the field (Fig.~\ref{fig:compare_mag}
and \ref{fig:compare_mag_perp}) as well as the magnetic energy density in
Fig.~\ref{fig:compare_Bsquared}.  Note that we only show a Taylor expansion of
the highly complex exact solution as derived in Appendix~\ref{sec:analytics}.
Strictly, this solution applies only near the sphere with an accuracy to
$\mathcal{O}((r-R)^{3/2})$ as well as for flow lines that have small impact
parameters initially at infinity. Using a different expansion, we verified that
the general solution has the appropriate behavior of the homogeneous magnetic
field at infinity in the upper half-space pointing towards the positive
$y$-coordinate axis, i.e.{\ }rightwards in Fig.~\ref{fig:compare_mag}. As
expected, the $y$-component of the magnetic field increases as we approach the
sphere since the field lines are moving closer to each other. In the immediate
vicinity of the sphere, the $\B$ field attains a dipolar $z$-component as the
field lines are carried around the sphere with the fluid and causes them to
bend in reaction to the ram pressure of the sphere.  As pointed out by
\citet{1980Ge&Ae..19..671B} the magnetic lines of force that end at the
stagnation point are strongly elongated as the swipe around the sphere parallel
to the line of flow reaching from the stagnation point into the rear.  This
leads to the unphysical increase of the magnetic field as it approaches the
line of symmetry in the wake and eventually to a logarithmic divergence of the
magnetic energy density there.

In the upper half-plane, the analytic solution matches the numerical one
closely.  Interestingly, in the region behind the deformed projectile, a {\em
  magnetic draping cone} develops that stems from the dynamically important
draping layer that has swiped around the sphere and advected downstream the
projectile. In addition, the magnetic pressure in the wake of the projectile is
also amplified by a moderate factor of roughly five
(cf.~Fig.~\ref{fig:compare_Bsquared}).  We will show further down, that this
field is generated together with vorticity in the draping layer.  In the {\em
  parallel plane} to the initial magnetic field, the magnetic draping cone
causes the stationary flow not to converge towards the symmetry axis and
protects the region in the wake against the increase of the magnetic energy
without bounds.  The numerical solution can qualitatively be obtained by
remapping the analytic solution for $\theta > \pi/2$ onto the coordinate along
the magnetic draping cone. In the {\em perpendicular plane} to the initial
magnetic field, there is even better agreement between the analytic and the
numerical solution. The magnetic field in that plane lies primarily in its
initial $y$-direction. This behaviour can easily be understood in terms of the
field lines sweeping around the sphere in a laminar flow. Numerically, we
simulate the response of the geometry of the projectile to the
hydrodynamics. Vortices in the wake deform the projectile leading to a
cap-geometry and a mushroom shape of the $y$-component of the magnetic
field. This implies that the flow lines detach from the dense material of the
projectile generating furthermore vorticity and MHD turbulence in the wake.
The turbulent field mixes the Cartesian components which can be nicely seen in
the Fig.~\ref{fig:compare_mag_perp}. The magnetic pressure summarizes our
results nicely showing the draping cone in the parallel plane and the mushroom
shaped magnetic energy density in the plane perpendicular to that
(cf.~Fig.~\ref{fig:compare_RhoVsquared}). Note that we choose the same color
scale as derived from the simulations which leads to a saturation of the
magnetic energy density in the kinematic approximation at the contact of the
spherical body and on the axis in the wake.

\section{CHARACTERISTICS OF MAGNETIC DRAPING}
\label{sec:characteristics}
\subsection{Field strength in draping layer}
\label{sec:strength}

\begin{figure}
\centering
\plotone{\choosefigure{magneticpressure.eps} {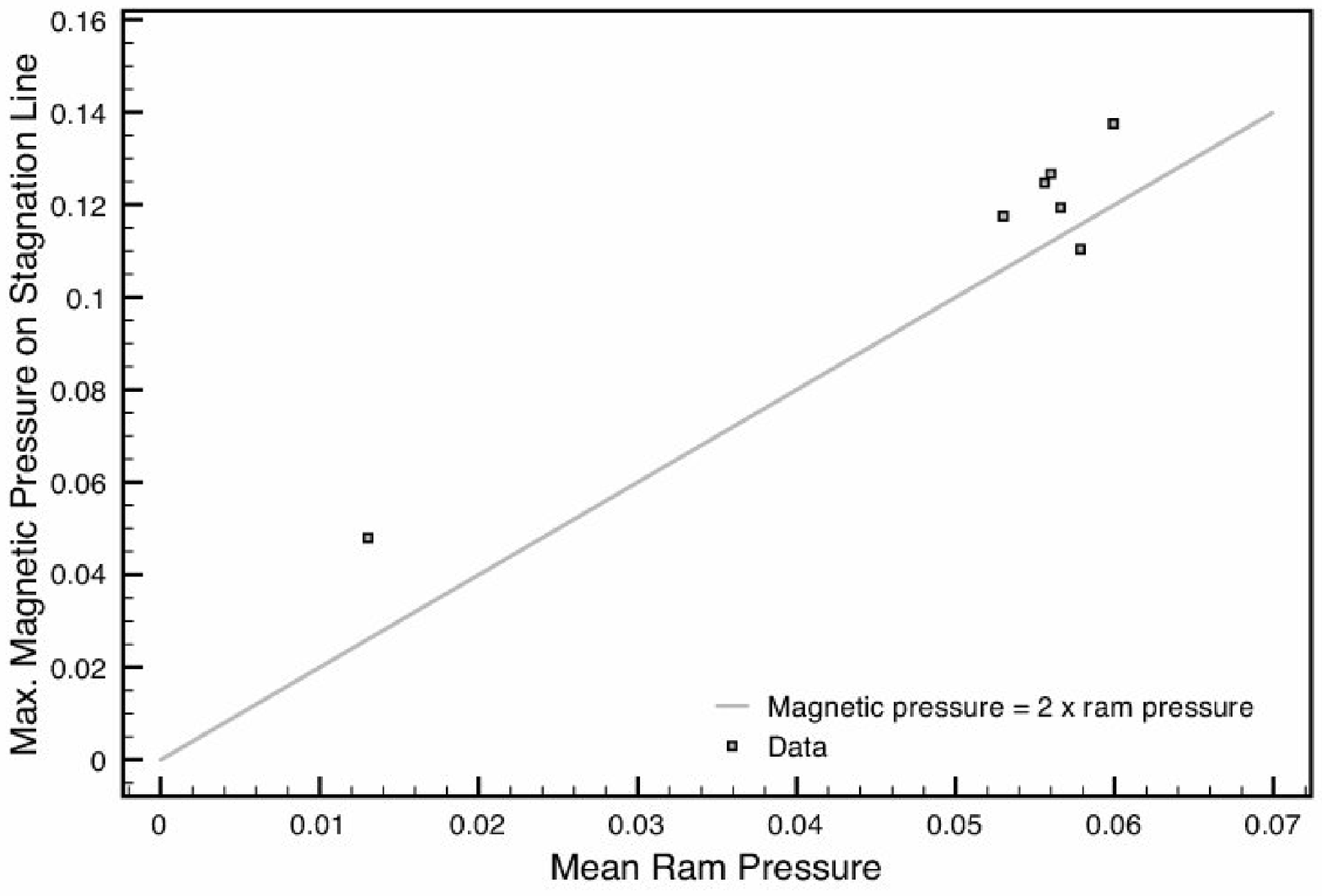} {magneticpressure.eps}}
\caption{A plot for the 3d runs presented here showing the magnetic
pressure on the stagnation line once a steady value had been achieved
for this quantity as a function of the mean ram pressure ($\rho \bra u\ket ^2$) as seen by
the projectile.   Omitted is run G, for which the magnetic layer was under-resolved
and thus the maximum magnetic field strength in the layer falls much lower; however,
as we will see, even this under-resolution does not strongly effect other global 
properties of the magnetic drape. \forhighresfiguressee}
\label{fig:magneticpressure-scatterplot}
\end{figure}

The kinematic solution predicts the magnetic pressure diverges at the
stagnation point, which is clearly unphysical.  From our discussions
in \S\ref{sec:picture} and \S\ref{sec:comparison}, we expect that the
magnetic pressure in the draping layer should be on order $\rho u^2$,
at which point the magnetic back-reaction begins to strongly effect
the flow; to first order there is no dependence on other
parameters, such as background magnetic field.  One would expect, too,
from looking at figures such as Fig.~\ref{fig:mag/ram-pressure} that
the maximum magnetic pressure should exceed the ram pressure by some
factor, as the magnetic pressure distribution at the head of the drape
is responsible for redirecting the flow in the plane of the draping.

We can test this by plotting, for all our runs, the steady maximum
magnetic pressure at the stagnation line (the field quantity that is
easiest to consistently characterize) versus the mean ram pressure
seen by the projectile, $\rho \bra u\ket ^2$, where $\bra u\ket $ is
the mean of the projectile velocity (calculated as in by
Eqn.~\ref{eq:bulletvelocity}) during the run, and $\rho$ is the ambient
density.  The plot is shown in
Fig.~\ref{fig:magneticpressure-scatterplot} and verifies our
expectation.

\subsection{Opening Angle}

\begin{figure}
\centering
\includegraphics[angle=-90,width=\bigfigsize]{\choosefigure {bubblewake-tubes-mag-morelines-halfvel.eps.epsf} {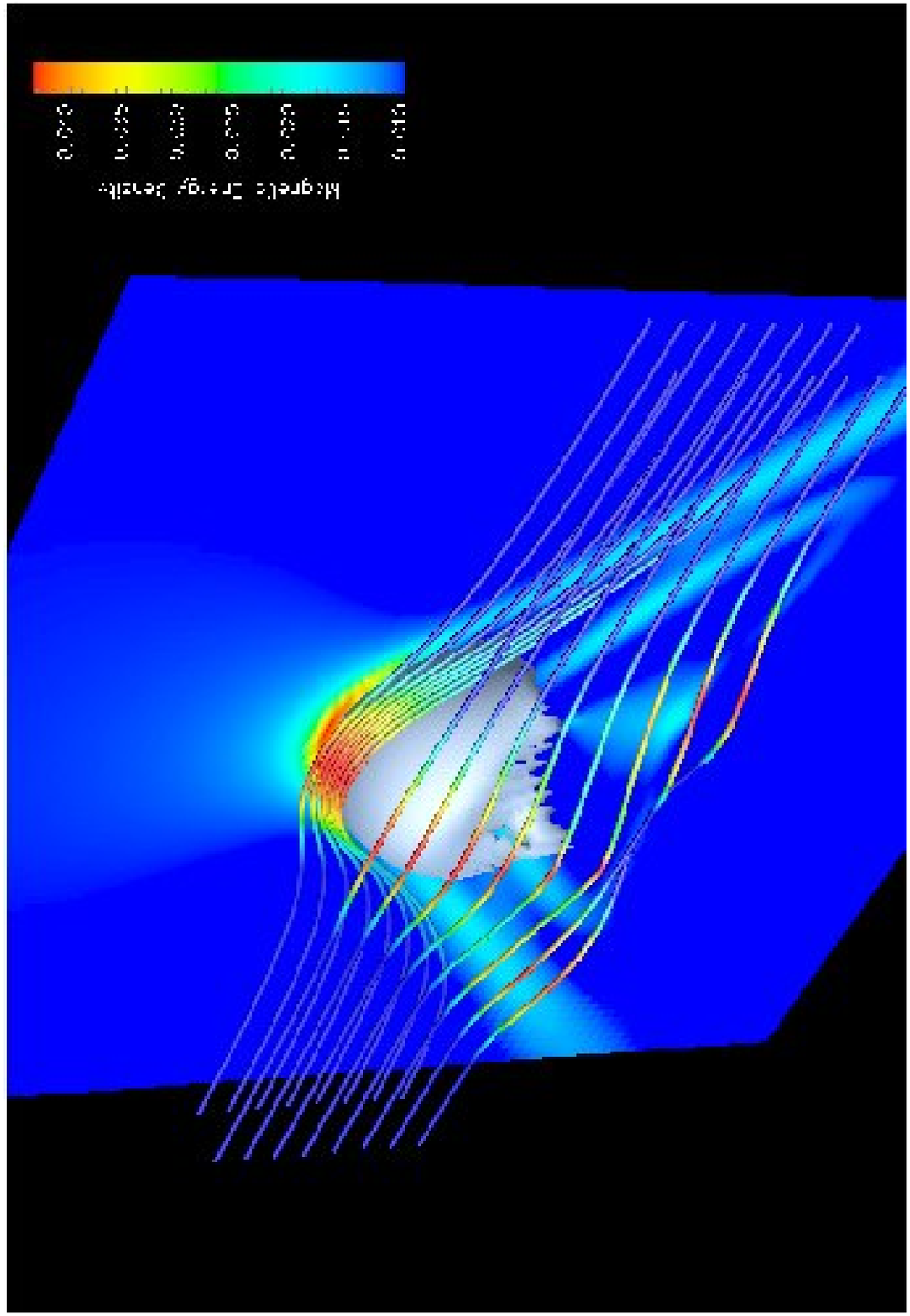} {bubblewake-tubes-mag-morelines-halfvel-cmyk.eps}}
\caption{As in Fig.~\ref{fig:3drendering-normalvel}, but for Run E; that is, with the projectile's
velocity reduced by a factor of one-half (so that $u = 0.125$ in code units).  \forthreedfiguressee}
\label{fig:3drendering-halfvel}
\end{figure}

The magnetic bow wave behind the projectile is expected to propagate
transversely away from the projectile at $\upsilon_A$ along the field
lines, and of course to fall behind the projectile at velocity $u$.
This suggests a natural opening angle in the plane along the magnetic
field, $\tan \theta = \upsilon_A / u$.   That the direction of the scaling
is correct can be determined by qualitative inspection of a sequence
of 3d renderings of simulation outputs as the velocity changes; e.g.,
Figs.~\ref{fig:3drendering-halfvel},\ref{fig:3drendering-normalvel},\ref{fig:3drendering-doublevel}
for $u = 1/8, 1/4, 1/2$ and $\upsilon_A$ fixed at $0.1414$.    

Although the field lines are stretched during the draping, it is the
initial $\vel_A$ that is relevant, as the stretching of the field
lines in the $z$-direction do not effect the propagation speed in the
$y$-direction.  For instance, consider a $\bf{\hat{z}}$-velocity shear
in $y$, $\vel = (0,0,y/\tau)$, with $\B = (0,B_0,0)$.  The induction
equation gives us $\dot{\B} = \nabla \times (\vel \times \B) =
(0,0,B_0/\tau)$, so that the magnetic field is only changed in the
$\bf{\hat{z}}$-direction; thus $\Alfvenvelmag_y = \Alfvenvel \cdot
\hat{y} = \Alfvenvelmag \hat{\B} \cdot \hat{y} = (|B|/\sqrt{4 \pi
  \rho}) (\B \cdot \hat{y})/|B| = B_y / \sqrt{4 \pi \rho} =
\Alfvenvelmag_{,0}$

\ifincludethreed
\begin{figure}[h!]
\centering
\includemovie[poster,3Dcoo=79.5 75 79.5, 3Droo=434.3960784034515, 3Dc2c=0 1 0]{5in}{5in}{3d/halfvel-small-compressed.u3d}
\caption{Interactive 3D version of Figure \ref{fig:3drendering-halfvel} above.}
\end{figure}
\fi

\begin{figure}
\centering
\includegraphics[angle=-90,width=\bigfigsize]{\choosefigure{bubblewake-tubes-mag-morelines-doublevel.eps.epsf} {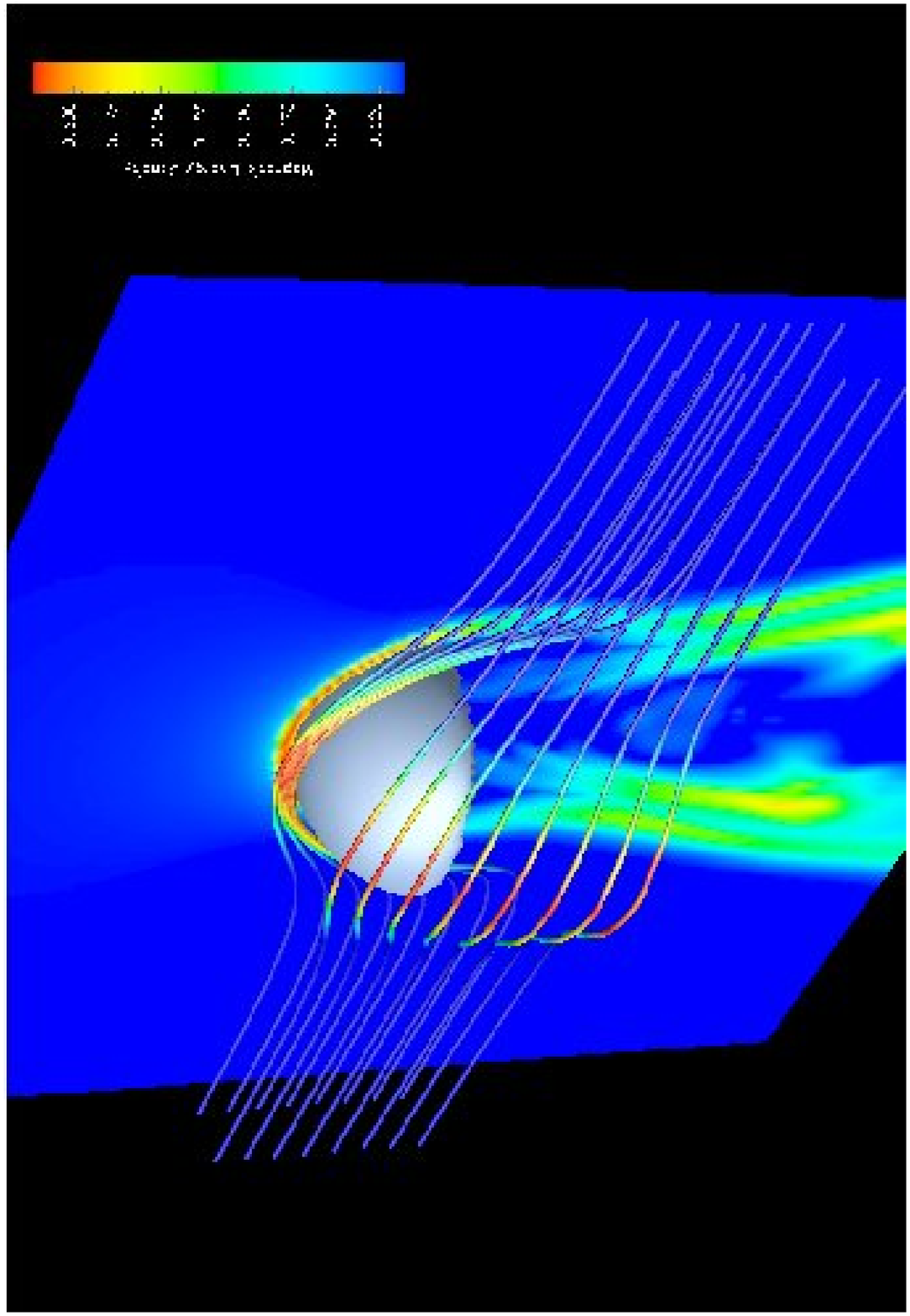} {bubblewake-tubes-mag-morelines-doublevel-cmyk.eps}}
\caption{As in Fig.~\ref{fig:3drendering-normalvel}, but for run G; that is, with the projectile's
velocity increased by a factor of two (so that $u = 0.5$ in code units).  \forthreedfiguressee}
\label{fig:3drendering-doublevel}
\end{figure}

\ifincludethreed
\begin{figure}[h!]
\centering
\includemovie[poster,3Dcoo=79.5 75 79.5, 3Droo=434.3960784034515, 3Dc2c=0 1 0]{5in}{5in}{3d/doublevel-small-compressed.u3d}
\caption{Interactive 3D version of Figure \ref{fig:3drendering-doublevel} above.}
\end{figure}
\fi

\begin{figure}
\centering
\plotone{\choosefigure{anglefigs.eps} {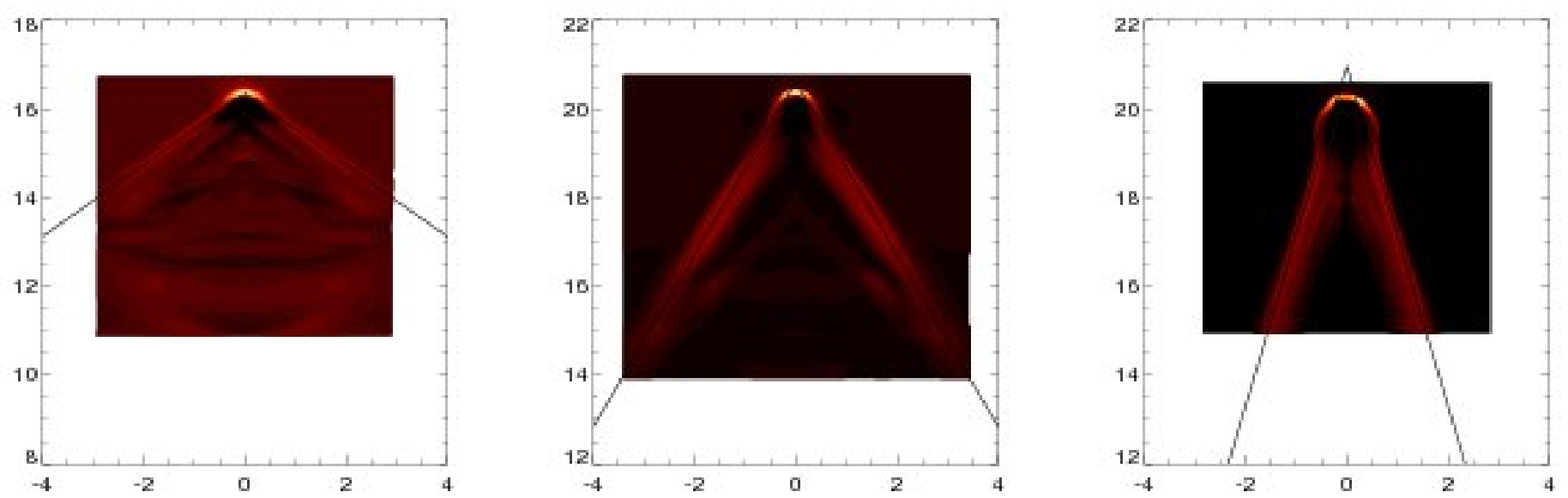} {anglefigs-bw.eps}}
\caption{Plot of magnetic energy density in the $y$,$z$ plane for simulations with $R = 0.5$ and,
left to right, $u = 0.125, 0.25, 0.5$; shown with black lines are the fitted opening angles of the
magnetic draping layer, omitting the region including the material from the projectile.   
The fit slopes (\eg{}, $\tan \theta$) are 1.24, 0.515, 0.261, and those predicted by
${\upsilon_A}_0/\bra u\ket $ are 1.13, 0.566, 0.283; this agreement is within 10\%.   \forhighresfiguressee}
\label{fig:openingangles}
\end{figure}

\begin{figure}
\centering
\plotone{\choosefigure{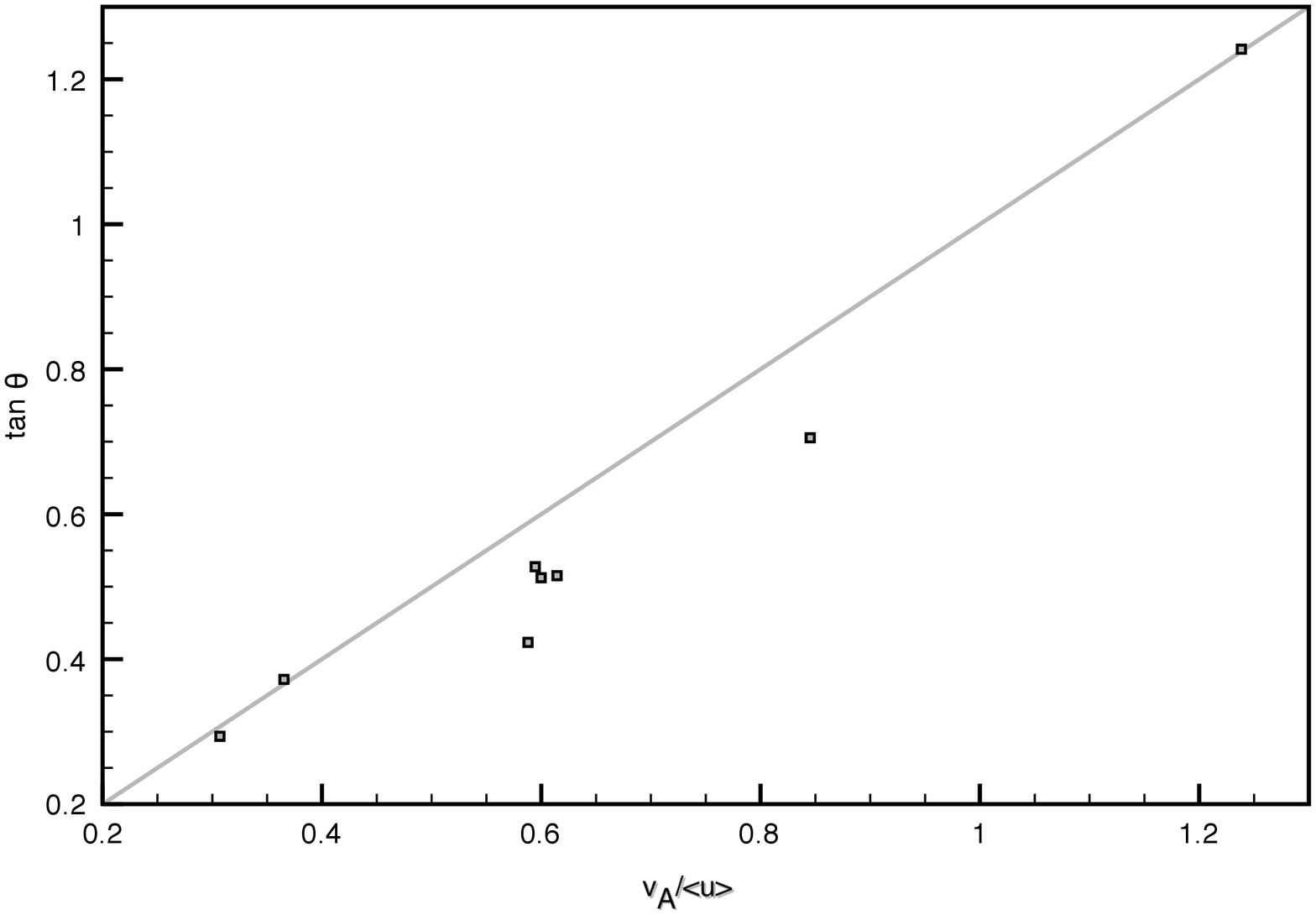} {openingangles-scatter.eps} {openingangles-scatter.eps}}
\caption{A plot for the 3d runs presented here showing the tangent of the fit opening angles
of the drape in the $yz$ plane versus $v_A/\bra u\ket $.   Data from all runs are shown.}
\label{fig:openingangles-scatterplot}
\end{figure}

One can quantify the agreement with this scaling by measuring the opening
angle for the drapes in our simulations.   The maxima of magnetic field
on either side of the stagnation line in the $y-z$ plane are found and
tabulated along the $z$ direction of the simulations, and -- omitting
the regions above or near the projectile itself, and the region below which 
the drape becomes weaker than transient features in the wake -- lines are fit,
and the slope gives the (half-)opening angle.   The results of the
fitting procedure are shown for the same three simulations in Fig.~\ref{fig:openingangles},
and a scatter plot for all are runs are given in Fig.~\ref{fig:openingangles-scatterplot}.
The scatter for this quantity, and agreement with the prediction, is somewhat worse
than for the other quantities we consider, possibly because the large-scale 
geometry of the draping is more sensitive to the boundaries and the finite size
of the domain than other, more local, quantities.

\subsection{Deceleration by magnetic tension}
\label{sec:deceleration}

\begin{figure}
\centering
\plotone{\choosefigure{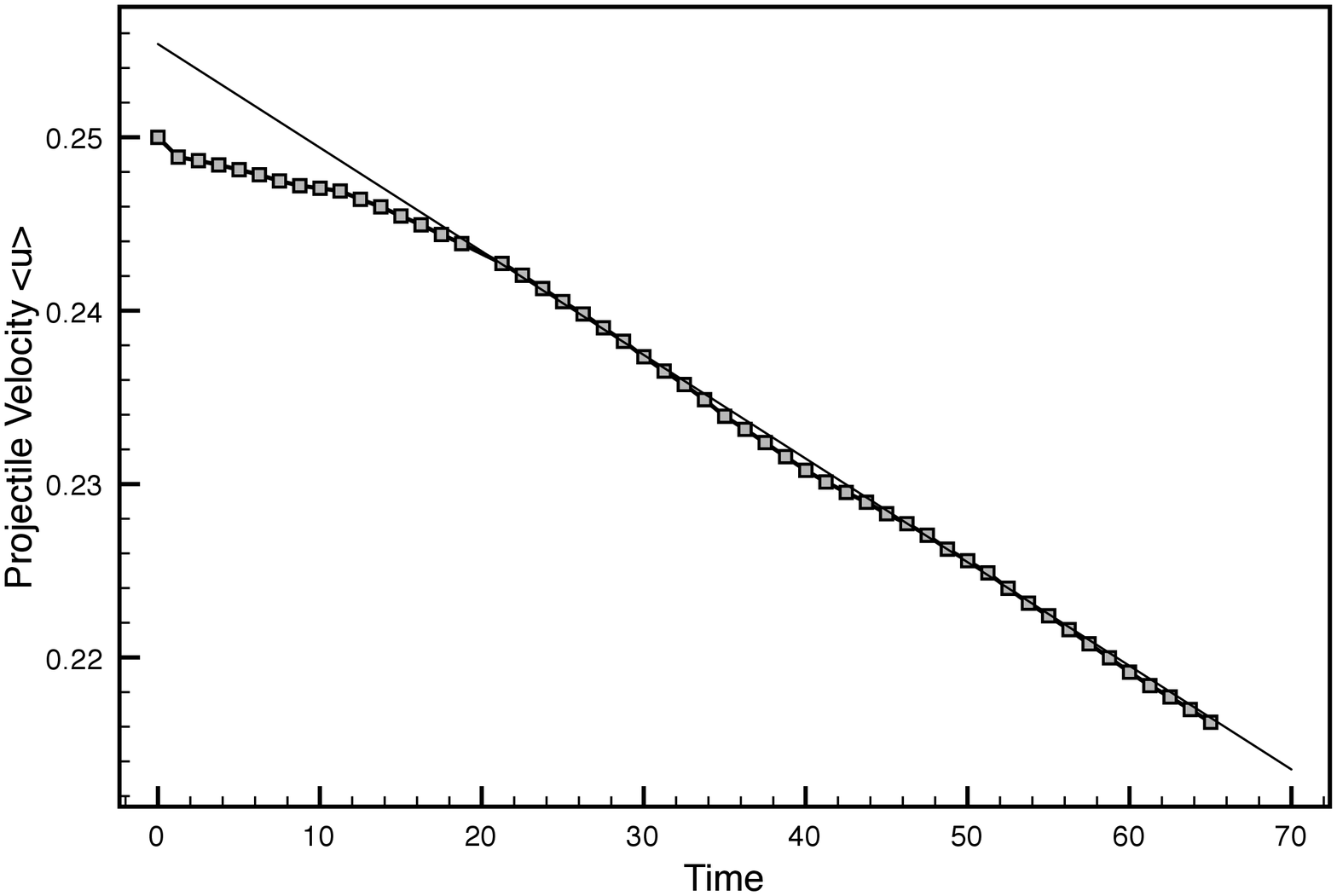} {decel-r05.eps} {decel-r05.eps}}
\caption{A plot showing, with squares, the evolution of projectile
velocity (calculated as in by Eqn.\ref{eq:bulletvelocity}) over time for
run F.   Note that the projectile encounters the magnetic field at time 20 
in these units.  Plotted as a thin line is the best fit deceleration,
$\dot u = -5.98 \times 10^{-4}$, for times greater than 20.}
\label{fig:decelleration-r05}
\end{figure}

In the scatter plots presented above, we use the mean velocity
$\bra u\ket$ of the projectile over time, because there is a measurable
deceleration of the projectile.   An example, for run F, is shown in
Fig.~\ref{fig:decelleration-r05}.   Before the projectile encounters
the magnetic field at $z = 10$, hydrodynamic drag -- in principle either
(numerical) viscous drag or the the drag force caused by the creation
of a turbulent wake -- is all that can play a role, and for the simulations
presented here, it is the second which dominates.   The well-known form
for the drag on a sphere is $F_\rmn{D} = 1/2 \rho u^2 A C_\rmn{D}$, or in terms of a
deceleration,
\begin{equation}
\dot u_\rmn{D} = -\frac{3}{8} \frac{\rho u^2}{\bra\rho_\rmn{p}\ket R} C_\rmn{D}
\label{eq:hydrodecel}
\end{equation} 
where $C_\rmn{D}$ is the drag coefficient, experimentally known to be
between $0.07 - 0.5$, with $0.5$ for a turbulent wake, $\rho_\rmn{p}$ is the
density and $A$ is the cross-sectional area of the projectile in the
direction of motion.

\begin{figure}
\centering
\plotone{\choosefigure{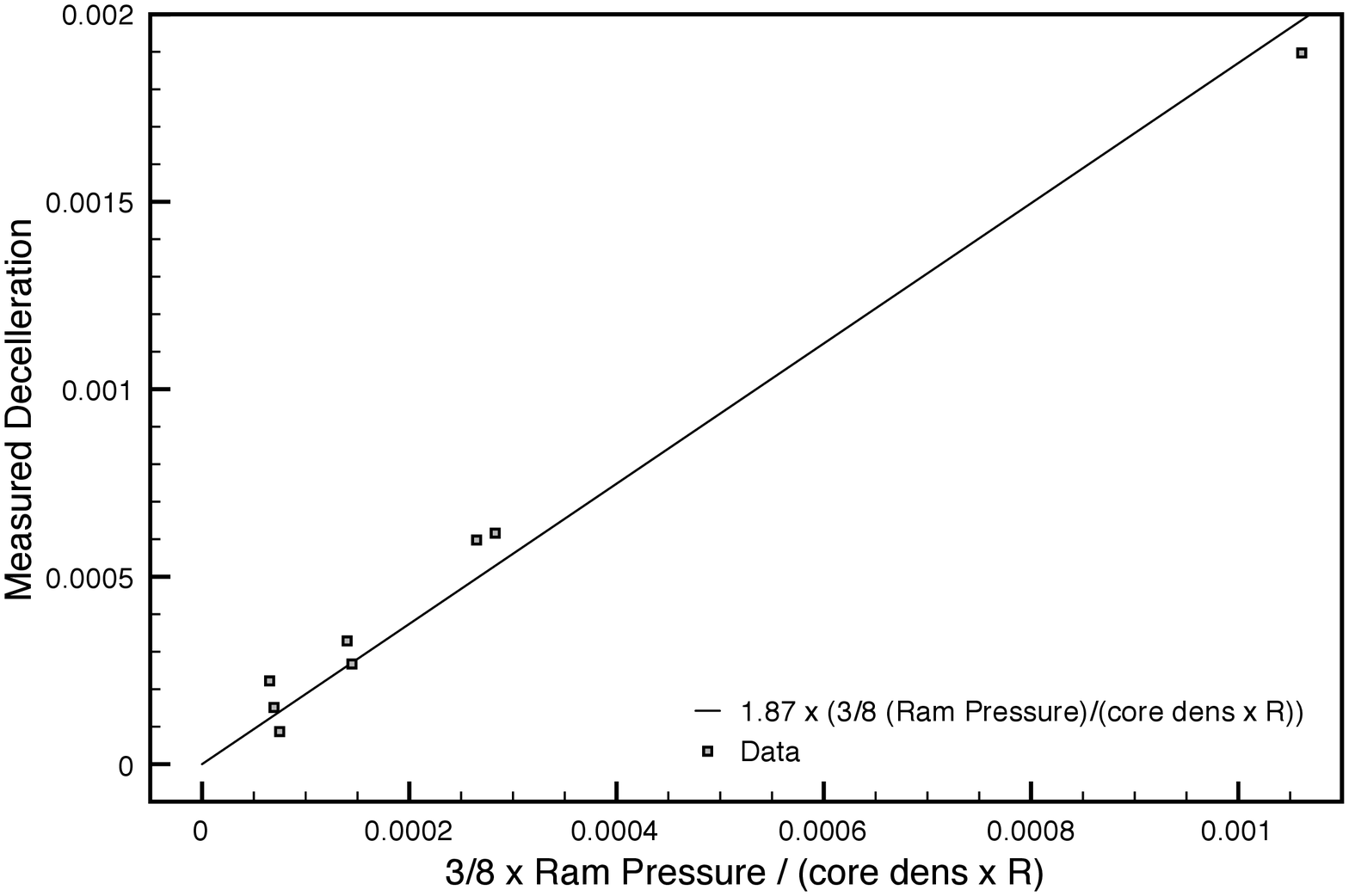} {deceleration.eps} {deceleration.eps}}
\caption{A plot for the 3d runs presented here showing the measured
deceleration of the projectile versus the functional form we expect it to
take, proportional to $3/8 \rho u^2/ (\bra \rho_\rmn{p} \ket R)$, where $\bra
\rho_\rmn{p} \ket$ is the mean density of the projectile (in code units, $\approx 150$).
Because the magnetic field strength and curvature varies over the draped layer, there is
an undetermined geometrical factor in the magnitude of the deceleration;
we find it here to be approximately $1.87$.  }
\label{fig:decelleration-scatterplot}
\end{figure}

However, once the magnetized region is reached and a magnetic layer
built up, then another force acts on the projectile -- the magnetic
tension from the stretched field-lines.  This transition can be seen in
Fig.~\ref{fig:decelleration-r05} for run F; other runs behave similarly.
We see that the deceleration caused by the magnetic field draping
is actually significantly stronger than the hydrodynamic draping.
This magnetic tension force is $F_\rmn{T} = B^2/(4 \pi R)$; we know the magnetic
strength in the draping layer scales as $\rho u^2$ (\S\ref{sec:strength})
and so we can write the deceleration as
\begin{equation}
\dot u_\rmn{T} = -\frac{3}{8}\frac{\rho u^2}{\bra\rho_\rmn{p}\ket R} C_\rmn{G}
\label{eq:magneticdecel}
\end{equation} 
where $C_\rmn{G}$ is a geometric term taking into account the fact
that both the magnetic field strength and radius of curvature of the
field lines vary over the `cap' of the projectile, and we have chosen
to normalize $C_\rmn{G}$ so that Eqns. \ref{eq:hydrodecel} and
\ref{eq:magneticdecel} have the same numeric prefactor for convenience
in comparison.  We can test this scaling, and at the same time
empirically obtain $C_\rmn{G}$, by plotting the decelerations for our
different runs, as is done in
Fig.~\ref{fig:decelleration-scatterplot}; we find $C_\rmn{G} \approx 1.87$.

It is interesting to note first that the two deceleration terms scale
in the same way, so that their relative importance remains constant;
and that said ordering is such that the magnetic tension deceleration
is always more important, by a factor of $\approx 3.7$, for the case
of highly turbulent ($\Reynolds \approx 1000$) hydrodynamic drag of
$C_\rmn{D} = 0.5$.    In the case of our simulations, we do not
have the resolution to achieve that highly turbulent state.   The
effective Reynolds number of our simulations can be estimated by
examining the hydrodynamic drag, for example in the first 20 time
units of Fig.~\ref{fig:decelleration-scatterplot}.   This does not
quite give enough data to make a good reading, so we ran four
simulations with the fiducial parameters ($R = 1, \Delta x/R = 32$),
varying $u$, ($0.125,0.25,0.5,0.75$) and outputting only $u(t)$.
An excellent fit to the data is provided by $C_\rmn{D} \approx 0.77$,
which corresponds to (see, \eg{}, Fig.~34 in IV,\S45 of \cite{1959flme.book.....L})
a $\Reynolds$ of just under $200$; even in this more viscous case, 
the magnetic draping deceleration exceeds the hydrodynamic deceleration
by a factor of 2.5.

\subsection{Vorticity generation}
\label{sec:vorticity}

\begin{figure}
\epsscale{.8}
\centering
\plotone{\choosefigure{streamlines-r2.eps.epsf} {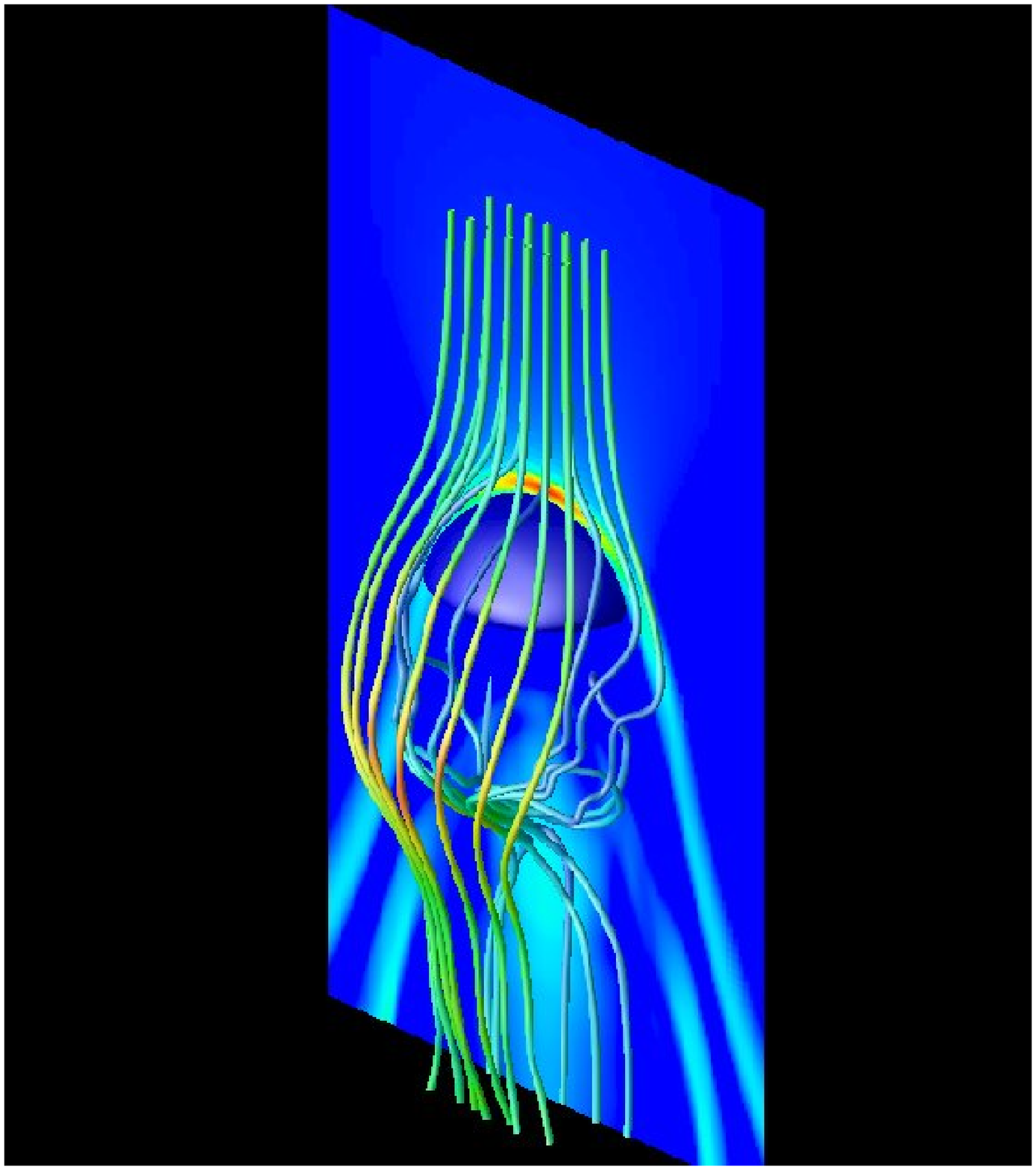} {streamlines-r2-cmyk.eps}}
\epsscale{1}
\caption{Plot of streamlines over the $R = 2$ projectile through a $\beta = 100$ medium.   Streamlines
are calculated in the frame of the mean velocity of the projectile.   The streamlines are coloured by
the magnitude of velocity, and the plane is once again colored by magnetic energy density.   At this time,
no instabilities have developed in the plane perpendicular to the ambient magnetic field, so fluid flows
smoothly over the projectile in this plane; however, fluid traveling close to the other plane experience
a gain of vorticity.  \forthreedfiguressee}
\label{fig:streamlines}
\end{figure}

\ifincludethreed
\begin{figure}[h!]
\centering
\includemovie[poster,3Dcoo=79.5 75 140, 3Droo=434.3960784034515, 3Dc2c=0 1 0]{5in}{5in}{3d/streamlines-small-compressed.u3d}
\caption{Interactive 3D version of Figure \ref{fig:streamlines} above.}
\end{figure}
\fi

\begin{figure}
\centering
\plotone{\choosefigure{vort-rcore2-90.eps} {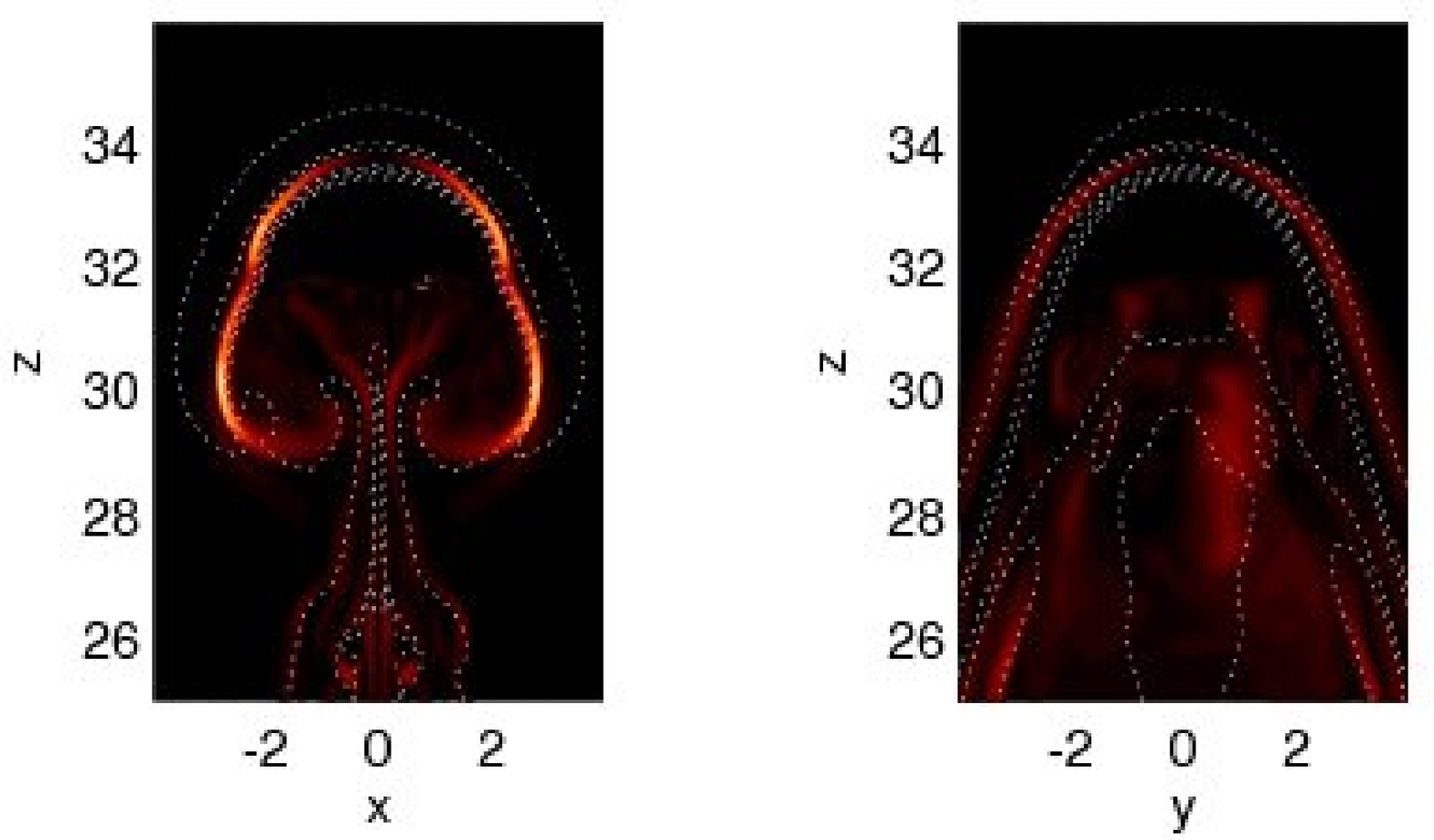} {vort-rcore2-90-bw.eps}}
\caption{The magnitude of vorticity is shown in color in the plane across and
  along to the initial magnetic field (left and right panel). The dotted
  lines represent contours of magnetic energy density.  \forhighresfiguressee}
\label{fig:vorticity}
\end{figure}

\begin{figure}
\centering
\plotone{\choosefigure{vortterms-rcore2-90.eps} {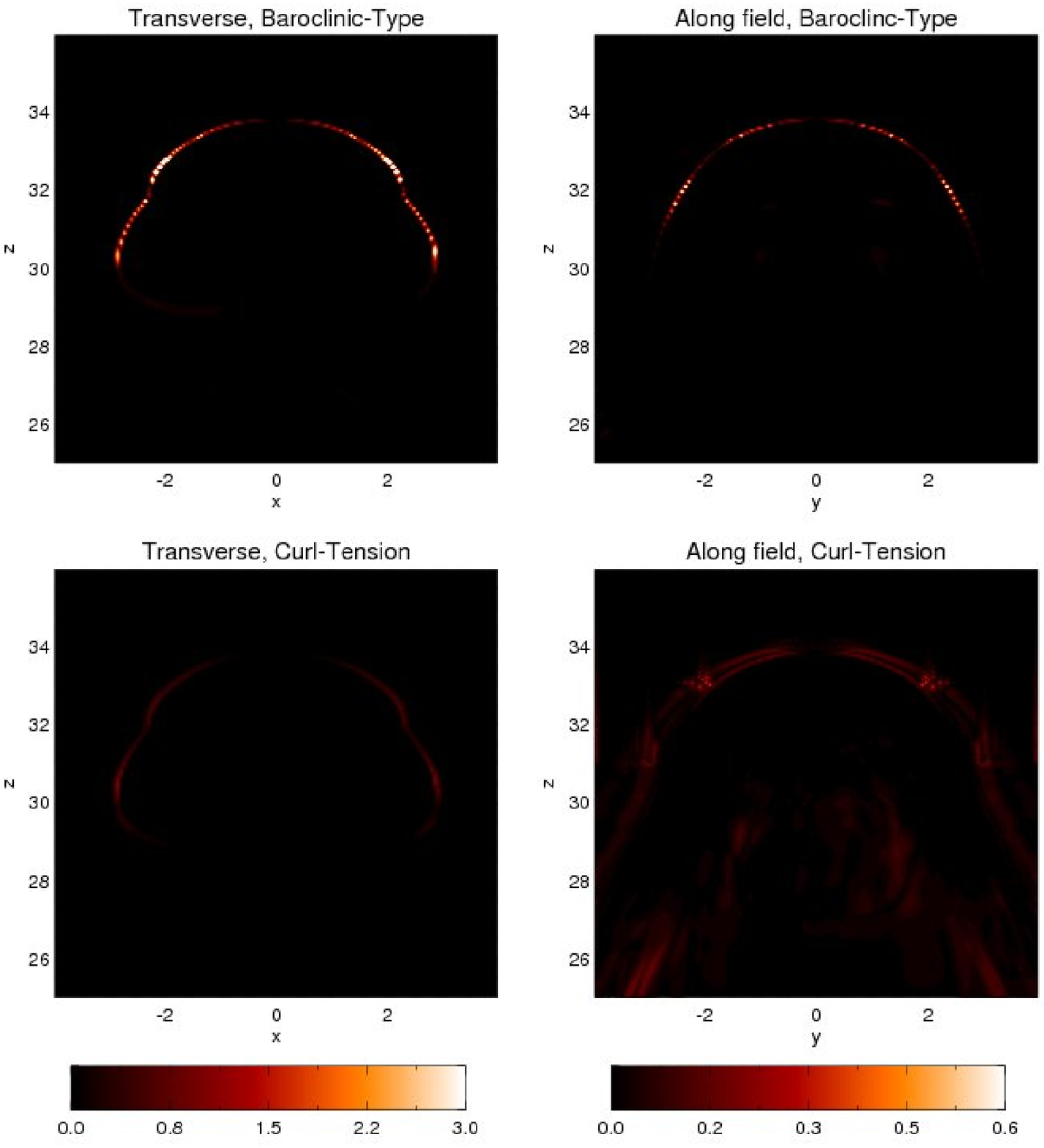} {vortterms-rcore2-90-bw.eps}}
\caption{Plot of the source terms for the specific vorticity injection rate as
  defined in Eqn.~\ref{eq:flux_freezing} for the plane transverse/parallel
  to the initial magnetic field (left/right panels).  The solid white
  contour shows the boundary between projectile material and ambiemt
  material.  The upper panels show the contribution of the baroclinic-type
  term where $\nabla \rho$ is not aligned with the thermal and magnetic
  pressure force.  Due to the large density gradient, this term dominates
  the vorticity injection at the interface between the projectile and
  the ambient medium. The curl of the magnetic tension force seems to
  be the dominant injection mechanism in the wake.  Note the different
  (linear) colour scales for the planes across and along the initial
  magnetic field.  \forhighresfiguressee}
\label{fig:vorticity_sources} 
\end{figure}

The flow pattern around a moving body looks as follows for large Reynolds
numbers. The flow is laminar and reflects a potential flow solution in almost
the entire volume except for a narrow boundary layer and the turbulent
wake. The characteristic drag coefficient decreases as the Reynolds number
increases \citep{1959flme.book.....L}.  This turbulent wake exerts a drag on
the body that decelerates it as described in \S\ref{sec:deceleration} and shown
in Fig.~\ref{fig:decelleration-r05}. This figure shows two distinctive
deceleration regimes where only the turbulent drag is present in the initial
phase, where the magnetic field has not been switched on, and a magnetic
tension dominated drag phase at later times. 

An independent argument is provided by Fig.~\ref{fig:streamlines}.  In the pure
hydrodynamic case, we do not expect any statistical anisotropy of the flow
pattern around the moving body. However in our MHD flow, there is an
unambiguous anisotropy visible for the stream lines. In the plane perpendicular
to the ambient initial magnetic field, the fluid flows smoothly over the
projectile with only mild perturbations for streamlines near the boundary
layer. In the plane of the initial magnetic field where the draping cone forms,
the stream lines are bend towards the turbulent wake and experience the
generation of vorticity $\bomega = \nabla\times\vel$. The magnitude of
vorticity in our simulations is shown in Fig.~\ref{fig:vorticity}.  Vorticity
is generated as the fluid enters the region in the draping layer where magnetic
field lines are slipping around the projectile, in particular in the plane
transverse to the initial magnetic field. The resulting velocity field can not
any more be described by the potential flow solution which causes the
analytical solution to break down at the magnetic draping layer and behind the
magnetic shoulder. The vorticity in the wake suggests the presence of MHD
turbulence that might be responsible for stretching and amplifying the magnetic
field furthermore.

We are interested how exactly the topology of the magnetic draping layer can be
responsible for generating vorticity into an initially vorticity-free flow
pattern.  The equation of motion for an inviscid and magnetized fluid without
gravity may be written in the form
\iftwocol
\begin{eqnarray}
\label{eq:Euler}
\rho\,\frac{\dd \vel}{\dd t} &=& 
\rho\,\partf{\vel}{t} + \rho\,(\vel\cdot\nabla)\,\vel = 
-\nabla P + \j\times\B \nonumber\\
&=&
-\nabla \left(P +\frac{B^2}{8\pi}\right)
+\frac{1}{4\pi}(\B\cdot\nabla)\,\B,
\end{eqnarray}
\else
\begin{equation}
\label{eq:Euler}
\rho\,\frac{\dd \vel}{\dd t} = 
\rho\,\partf{\vel}{t} + \rho\,(\vel\cdot\nabla)\,\vel = 
-\nabla P + \j\times\B=
-\nabla \left(P +\frac{B^2}{8\pi}\right)
+\frac{1}{4\pi}(\B\cdot\nabla)\,\B,
\end{equation}
\fi
where we define the convective derivative in the first step and applied
$\nabla\times \B = 4\pi\,\j$ in the last step. The first term on the right-hand
side describes the potential force due to the sum of the isotropic thermal
pressure $P$ and magnetic pressure $B^2/(8\pi)$, while the second term
describes the magnetic tension force.  Applying the curl operator to
Eqn.~(\ref{eq:Euler}) and identifying the vorticity $\bomega =
\nabla\times\vel$, we arrive at the equation governing the evolution of
vorticity:
\iftwocol
\begin{eqnarray}
\label{eq:flux_freezing}
\frac{\dd }{\dd t}\left(\frac{\bomega}{\rho}\right) &=& 
\left(\frac{\bomega}{\rho}\cdot\nabla\right)\vel + 
\frac{1}{4\pi\,\rho^2}\,\nabla\times(\B\cdot\nabla)\,\B \nonumber\\
&+&
\frac{1}{\rho^3}\,\nabla\rho\times
\left[\nabla \left( P + \frac{B^2}{8\pi}\right) - 
\frac{1}{4\pi}\,(\B\cdot\nabla)\,\B\right].
\end{eqnarray}
\else
\begin{equation}
\label{eq:flux_freezing}
\frac{\dd }{\dd t}\left(\frac{\bomega}{\rho}\right) = 
\left(\frac{\bomega}{\rho}\cdot\nabla\right)\vel + 
\frac{1}{4\pi\,\rho^2}\,\nabla\times(\B\cdot\nabla)\,\B +
\frac{1}{\rho^3}\,\nabla\rho\times
\left[\nabla \left( P + \frac{B^2}{8\pi}\right) - 
\frac{1}{4\pi}\,(\B\cdot\nabla)\,\B\right].
\end{equation}
\fi
This equation describes the condition that the vorticity is `frozen' in the
plasma if the last two terms are negligible.\footnote{This can be seen by
  considering the evolution of an infinitesimal vector $\delta \x$ connecting
  two neighboring fluid parcels, as the fluid moves with the velocity
  field. The point initially at position $\x$ at time $t$ will be displaced to
  the position $\x+\vel(\x)\Delta t$ at time $t+\Delta t$. The neighboring
  point initially at $\x+\delta\x$ at time $t$ will be displaced to the
  position $\x+\delta\x+\vel(\x+\delta\x)\Delta t$ at time $t+\Delta t$. Hence
  this `frozen' connecting line evolves according to
\begin{equation}
\label{eq:x_flux_freezing}
\frac{\dd }{\dd t}\left(\delta\x\right) = 
(\delta\x\cdot\nabla)\vel.
\end{equation}
which resembles Eqn.~(\ref{eq:flux_freezing}) if we neglect the last two terms
and identify $\delta\x = \eps \bomega/\rho$ initially, where $\eps>0$ is a
small quantity. Since the differential equation is true for any time, the same
relation will hold for all times for the vorticity.} Vorticity is necessarily
generated, if the curl of the force field generated by magnetic tension does
not vanish (referred to as {\em curl-tension term}). Another source of
vorticity is given by a flow where $\nabla\rho$ is not aligned with the
potential force due to thermal or magnetic pressure as well as the magnetic
tension force (referred to as {\em baroclinic-type term}).
Figure~\ref{fig:vorticity_sources} studies the relative importance of both
source terms.  Due to the large density gradient that develops at the interface
between the projectile and the ambient medium, the baroclinic-type term
dominates the vorticity injection at this interface; but it is unimportant elsewhere. The curl of
the magnetic tension force seems to be the dominant injection mechanism in the
wake, and contributes over a broader spatial range in the magnetic draping layer. We caution the reader that we cannot quantify the level of vorticity
injected by means of a turbulent boundary layer and refer to our phenomenological
argument at the beginning of this section that clearly indicates the importance
of the magnetic draping layer for the vorticity injection.

\section{INSTABILITIES}
\label{sec:instabilities}
The magnetic tension force as well as the magnetic layer geometry has
implications for the instabilities experienced by the projectile.  In
\S\ref{sec:picture} and \S\ref{sec:characteristics}, we saw that the
flow in the plane parallel to the initial magnetic field is stable and
the hydrodynamic instabilities are suppressed by the magnetic draping
layer \citep[suggested by][]{lintheory}.  In contrast, the flow around
the projectile in the plane transverse to the initial magnetic field
is unstable to Kelvin-Helmholtz and Rayleigh-Taylor instabilities, not
stabilized by the presence of magnetic field. We will show that the
Kelvin-Helmholtz instability remains stronger and leads to gradual
disruption of the projectile, although the impact of the Rayleigh
Taylor instability in our MHD case is greater than the purely
hydrodynamical case because of the greater deceleration.  For an
homogeneous initial magnetic field the induced vorticity remains
largely two-dimensional.

\begin{figure}
\centering
\plotone{\choosefigure{overplotDensityVsquared.eps} {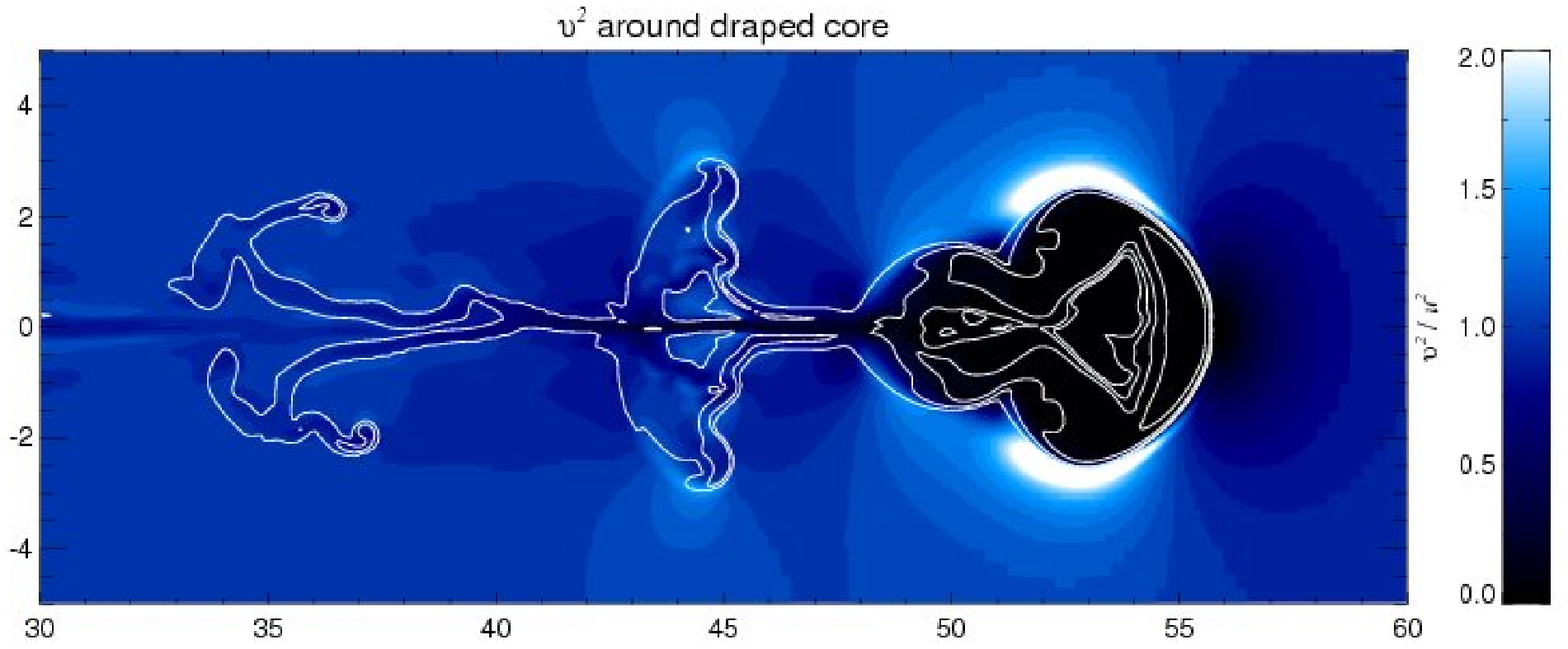} {overplotDensityVsquared.bw.eps}}
\caption{Kelvin-Helmholtz instability dissolves the projectile in the
  plane perpendicular to the initially homogeneous magnetic field. The
  flow is accelerated at the density enhancements of the stripped
  material due to the Bernoulli effect. In the wake of the projectile,
  there is a characteristic length scale of $\sim 10$ length units
  between the striped material which corresponds to an unstable mode
  with a wavelength of $2/3 R$.  \forhighresfiguressee}
\label{fig:Kelvin-Helmholtz}
\end{figure}

The projectile is being decelerated by magnetic tension as shown in
\S\ref{sec:deceleration}. This makes the projectile subject to {\em
  Rayleigh-Taylor instabilities} with the growth rate \citep{chandra}
\begin{equation}
  \label{eq:RT}
  \omega_\rmn{RT}^2 =  
  \frac{\bra\rho_\rmn{p}\ket-\rho_0}{\bra\rho_\rmn{p}\ket+\rho_0}\,\dot{u}_\rmn{T}\,k \simeq
  \frac{3}{8} \frac{2\pi\,\rho_0\,C_\rmn{G}}{\bra \rho_\rmn{p}\ket\,}\frac{u^2}{R^2}\frac{k}{k_0} \ge
  \frac{3\pi\,\rho_0\,C_\rmn{G}}{4 \bra \rho_\rmn{p}\ket\,}\frac{u^2}{R^2},
\end{equation}
where $k_0=2\pi/R$ defines the the smallest wavenumber of the system.
We also neglect viscosity and surface tension effects, and we work in the
limit where $\bra \rho_\rmn{p}\ket\gg \rho_0$.

The flow around the projectile causes a shear at the interface of the
projectile that can get non-linear by means of the {\em
  Kelvin-Helmholtz instability} and has the growth rate
\citep{chandra}
\begin{equation}
  \label{eq:KH}
  \omega_\rmn{KH} =  
  \frac{\sqrt{\bra\rho_\rmn{p}\ket\,\rho_0}}{\bra\rho_\rmn{p}\ket+\rho_0}\,\Delta u\,k 
  \simeq \frac{3\pi\,u}{R}\sqrt{\frac{\rho_0}{\bra\rho_\rmn{p}\ket}}\,\frac{k}{k_0}\ge
  \frac{3\pi\,u}{R}\sqrt{\frac{\rho_0}{\bra\rho_\rmn{p}\ket}}.  
\end{equation}
Here we neglect viscosity and self-gravity of the projectile and apply
the maximal velocity shear from the potential flow solution around a
spherical body, $\vel=3/2\, u\,\mathbf{e}_\theta$, which is valid at
$r=R$ and $\theta=\pi/2$. 

Which instability will eventually dominate and set the relevant
timescale? It turns out that the ratio of growth rates is independent
of the projectile properties and only depends on the wave number of
the considered mode,
\begin{equation}
  \label{eq:RT-KH}
  \frac{\omega_\rmn{KH}^2}{\omega_\rmn{RT}^2} \simeq 
    \frac{12\pi}{C_\rmn{G}}\,\frac{k}{k_0}\ge \frac{12 \pi}{C_\rmn{G}}\simeq 20.
\end{equation}
where from the previous section, $C_G \approx 1.87$ takes into
account the fact that both the magnetic field strength and radius of
curvature of the field lines vary over the `cap' of the projectile.
The largest length scale of the problem is given by the size of the
projectile in the direction of motion and sets the largest timescale
of the problem,
\begin{equation}
  \label{eq:RT-KH2}
  \frac{T_\rmn{KH}}{T_\rmn{RT}} \simeq \frac{1}{2}\sqrt{\frac{C_\rmn{G}\,k_0}{3 \pi\,k}}
  \le  0.22.
\end{equation}
Thus, we expect the Kelvin-Helmholtz instability in the plane
transverse to the initial magnetic field to be responsible for the
eventual disintegration of the projectile. These considerations allow us to
estimate the associated time- and length-scale on which we expect to
see the projectile material in the boundary layer to become unstable,
\begin{equation}
  \label{eq:L_KH}
  L_\rmn{KH} = T_\rmn{KH}\, u = \frac{2\pi\,u}{\omega_\rmn{KH}} \simeq
  \frac{2R}{3}\sqrt{\frac{\bra\rho_\rmn{p}\ket}{\rho_0}} \frac{k_0}{k}\le
  \frac{2R}{3}\sqrt{\frac{\bra\rho_\rmn{p}\ket}{\rho_0}}\simeq16.3
\end{equation}
in terms of the length units in the code.  This explains nicely the
instability features in the wake of Fig.~\ref{fig:Kelvin-Helmholtz}
that appear every 10 length units and indicate that a mode that is
slightly smaller than the projectile dimension is becoming unstable and
leads to a deposition of projectile material.

\section{DISCUSSION AND LIMITATIONS}
\label{sec:discussion}
We have investigated in detail the rapid formation of a magnetic draping
layer over a projectile, and examined some of the immediate dynamical
consequences.   It is worth considering how well these insights continue
to hold over longer timescales, and whether the draped field can offer
much protection over significant distances.

While details of how mixing might take place will depend sensitively
on the structure of the object in question, one requirement for a
projectile to mix significantly into the surrounding medium will be for
the projectile to sweep past on order its own mass in the ambient medium;
only then will there have been enough shear to significantly disrupt
the moving object.  This requires the projectile to traverse a distance
$L \sim (\bra \rho_\rmn{p} \ket /  \rho_0) R$.    For the runs considered in
previous sections, modeling this while continuing to resolve the magnetic
draping layer would require extremely costly simulations, even with AMR.

\begin{figure}[h!]
\centering
\includegraphics[angle=-90,width=\bigfigsize]{\choosefigure{dens75-mag.eps} {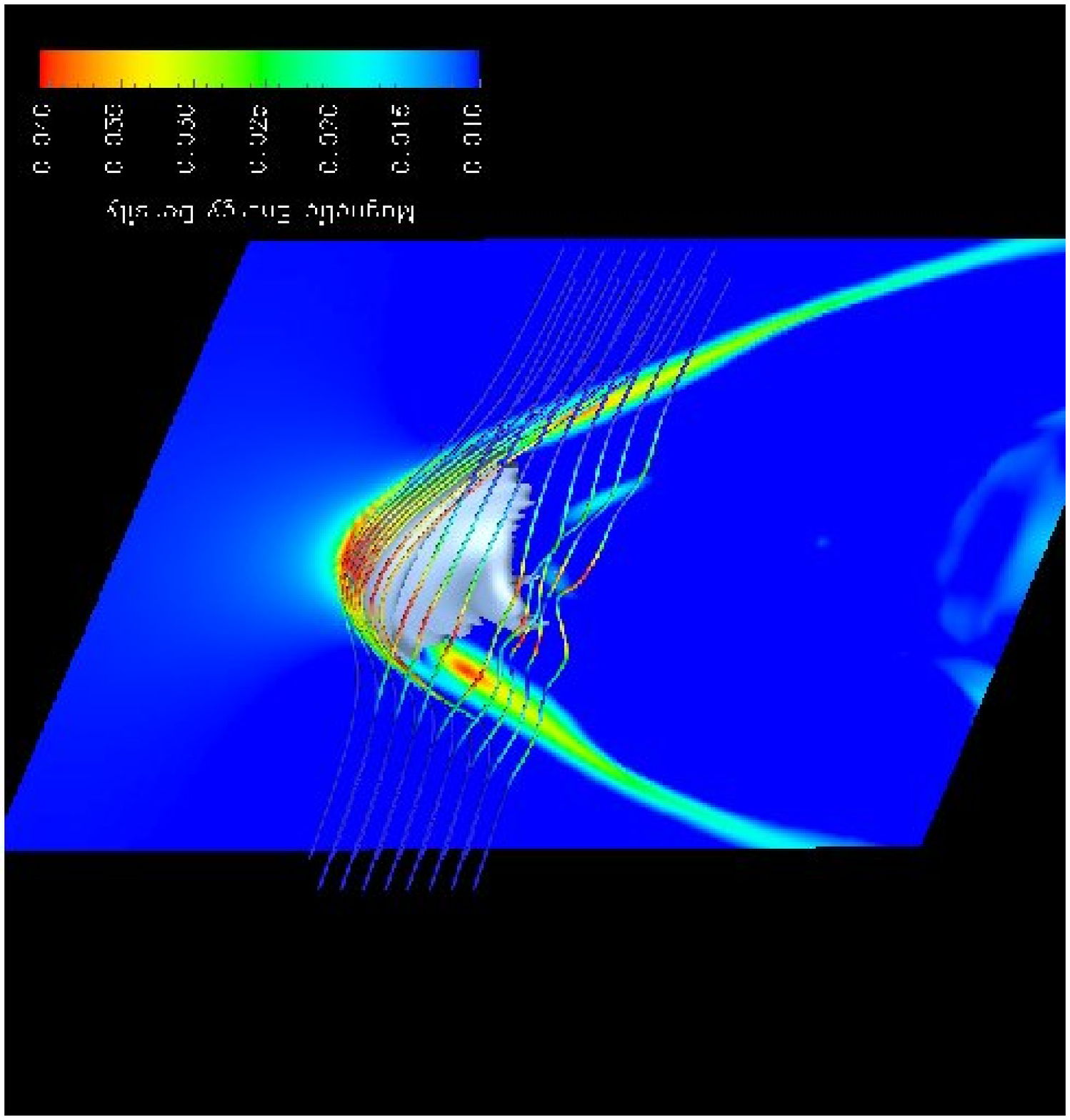} {dens75-mag-cmyk.eps}}
\caption{As in Fig.~\ref{fig:3drendering-normalvel}, but for the $\bra
\rho_\rmn{p} \ket = 15$ run, at a time where the projectile has swept past
approximately its own mass in ambient fluid.  \forthreedfiguressee}
\label{fig:3drendering-dens75}
\end{figure}

However, at the cost of complicating direct comparison with previous
simulations, one can gain some insight into what will happen over longer
times by considering those regions of parameter space which make the
computation more feasible.  In particular, for this section we perform an
analog to run B made with a maximum projectile density reduced by a factor of
10, so that $\bra \rho_\rmn{p} \ket/ \rho_0 \approx 15$.   With this reduced
density contrast, mixing happens more easily and the projectile sweeps
past its own mass in a computationally approachable time.   Results from
this run are shown in Fig~\ref{fig:3drendering-dens75}, at a time when
the projectile has approximately swept through its own mass of ambient medium.

In this run, the same features are seen as in previous sections; the
development of the strong narrow magnetic field layer, the opening angle
$\sim \Alfvenvelmag/u$, and the large-scale vorticity oriented primarily
along field lines generated in the wake.   However, over long times the
anisotropy imposed by the direction preferred by the magnetic field,
and as suggested in Fig.\ref{fig:drapinggeometry-cartoon}, becomes much
more pronounced, as the projectile becomes extremely aspherical; it is
greatly flattened along the direction of the magnetic field lines.

\begin{figure}
\centering
\plotone{\choosefigure{denscompare-magnomag.eps} {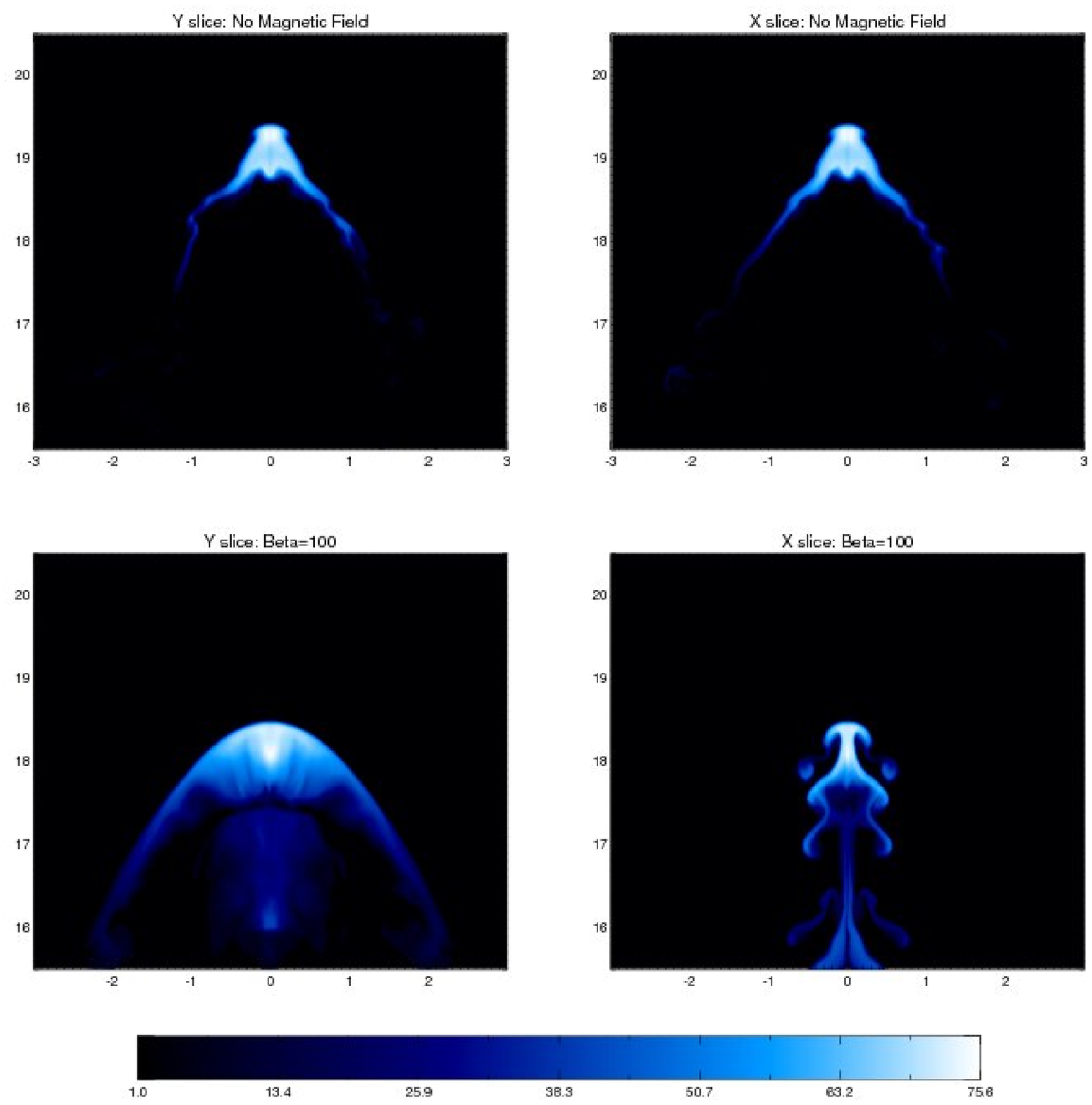} {denscompare-magnomag-bw.eps}}
\caption{Plots of density shown for the $\bra \rho_\rmn{p} \ket / \rho_0 = 15$
run, at a time when the projectile has swept past roughly its own mass of 
ambient medium.  Shown at top are simulations with no initial magnetic
field, and at bottom which a magnetic field as in Run B, with $\beta = 100$
and $\rho_0 u^2/P_{b,0} = 6.25$.  Panels on the left are along the $y$
direction (\eg{}, the direction of the initial magnetic field lines in
the second case) and along the $x$ direction (perpendicular to the magnetic
field lines) on the right.   These snapshots are taken at the same simulation
time for the two runs; the projectile in the magnetized case lags because of
the deceleration demonstrated in \S~\ref{sec:deceleration}.  \forhighresfiguressee}
\label{fig:dens75-mag-nomag}
\end{figure}

The long-time distortion of the projectile by the magnetic field -- which,
again, is initially strongly subthermal ($\beta \sim 100$) and a factor of
$6$ less than the ram pressure seen by the projectile -- is particularly
evident when seen compared to the results of the same projectile moving in
the absence of a magnetic field, as in Fig.~\ref{fig:dens75-mag-nomag}.
In this case, density is plotted in two cut planes at the same time
for the simulation with and without magnetic field.  \ifincludethreed
Three dimensional interactive density isosurfaces are also plotted in
Fig.~\ref{fig:dens75-mag-nomag-3d}.  \fi

Several features are immediately apparent.  The first is the significantly
different density distributions between the two scenarios.  The densest
material is more contained in the case with magnetic draping, but in the plane
along the magnetic field, material which is stripped off is more extended,
piling up along the draped magnetic contact.  The plane across the initial
magnetic field lines is even more interesting; here the stripped material is
much more contained, even with the presence of the Kelvin-Helmholtz instability.
Here stripped material stays almost completely within a cylinder of radius $R =
1$, the initial radius of the projectile, along the path of the projectile.

Also evident is that, although the two simulations are examined at the
same time, the projectile without magnetic fields is significantly further
ahead in the domain than the simulation with the magnetic field; this is
the result of the deceleration demonstrated in \S~\ref{sec:deceleration}.

\ifincludethreed
\begin{figure}[h!]
\centering
\includemovie[poster,3Dcoo=95.5 94 95.5, 3Droo=522., 3Dc2c=0 1 0]{3in}{3in}{3d/dens75-mag-compressed.u3d}
\includemovie[poster,3Dcoo=96 96 152.75, 3Droo=170, 3Dc2c=0 1 0]{3in}{3in}{3d/dens75-nomag-compressed.u3d}
\caption{Interactive 3D version of Figure \ref{fig:dens75-mag-nomag} above}.
\label{fig:dens75-mag-nomag-3d}
\end{figure}
\fi

In this work we have made several simplifying assumptions to allow us
to begin to understand the process of the draping.   We have neglected
consideration of the interior structure of the projectile, by for instance
omitting any self gravity which would be relevant for a minor merger.
This will effect the rate of stripping of material off of the core, and
thus long-term evolution, but is unlikely to directly effect the draping
process itself.    We have also not considered any gradient of properties
in the medium the projectile moves through; while this again would effect
long term behaviour, the set-up of the magnetic draping layer occurs so
quickly that it is unlikely that any background quantities would greatly
change over the small distances involved.

We have also omitted explicit treatment of dissipative microphysics 
\cite{lyutikovdissipation,schekochihindissipation}.  Following up
with simulations which included these effects self-consistency will be
important for examining in detail the resulting sharpness of the
cold fronts (as done, for example, by \cite{asai07}) and the different
dissipation physics may also well effect the long time behaviour of mixing.
However, the initial draping layer is set up, and its properties are
determined, on timescales much faster than the dissipative timescales,
so these results will be unaffected.

We have also considered here only subsonic motions through the ambient
medium.  Many of the astrophysical processes where draping is relevant
can be supersonic, and so an important next step is to consider this
case, where a bow shock will occur before the magnetic draping layer.
While the bow shock will almost certainly be well separated from the
magnetic draping layer, the shock will affect both the geometry of the
flow onto the draping layer and greatly amplify the importance of the
thermal pressure.  On the other hand, behind the bow shock the flow
will be subsonic, so much of the discussion here will directly apply.
Similar detailed studies of draping in the supersonic case will be
considered in future work.  Perhaps more seriously, we have considered
here only the simplest case of an initial magnetic field uniform over
the scales of interest. It will be necessary to consider more realistic
field geometries.  This, too, is being considered in future work,
and will require much more careful treatment of the detailed magnetic
structure of the field.

\section{CONCLUSION}
\label{sec:conclusion}

This work aims at understanding the morphology and the dynamical properties of
magnetic draping to set a solid ground for its astrophysical applications.  A
core, bullet, or bubble that moves super-alfv{\'e}nically in even a very weakly
magnetized plasma necessarily sweeps up enough magnetic field to build up a
dynamically important sheath around the object; the layer's strength is set by
a competition between `plowing up' of field and field lines slipping around the
core, and to first order depends only on the ram pressure seen by the moving
object.  This layer is developed very quickly, potentially faster than a
crossing time of the projectile. The energy density in the draped layer, at its
maximum, exceeds the ram pressure by a factor of two, necessary to
anisotropically redirect the flow.  This effect has important implications for
galaxy cluster physics as it suppresses hydrodynamic instabilities at the
interface of AGN bubbles. It naturally explains so-called `cold fronts' by
keeping temperature and density interfaces of merging cores sharp that would
otherwise be smoothed out by thermal conduction and diffusion. Other important
astrophysical implications of this effect include draping of the solar wind
magnetic field at the bow shock of the Earth as well as pulsar wind nebulae.

In this paper, we compare a simplified analytical solution of the problem that
neglects the back-reaction of the dynamically important magnetic field on the
potential flow with a high-resolution AMR simulation and find very good
agreement between both solutions in the region ahead of the bubble.  Non-linear
back-reaction of the magnetic field in the draping layer necessarily implies
the generation of vorticity in the flow. The induced vortices in the wake
deform the core hydrodynamically and eventually cause the magnetic sheath to
peel off. There is a strong indication that the vorticity generation is
responsible for the intermittent amplification and stretching of the magnetic
fields as well as the injection of MHD turbulence in the wake of the core. If
this withstands further critical analysis, this mechanism might have profound
astrophysical implications for the amplification and generation of
large-scale magnetic fields in the inter-galactic medium.

The magnetic layer, once fully developed, has a characteristic geometry which
we have shown here to be roughly conical in the plane along the magnetic field
lines, with opening angle $\theta \sim \arctan(\Alfvenvelmag/u)$, and remains
contained in the perpendicular plane, with the Kelvin-Helmholtz instability
acting on the object interface.  The magnetic tension in the layer
significantly decelerates the object, dominating over any hydrodynamic drag.

Over long times, the anisotropy imposed by the field -- despite the fact
that the field is initially highly subthermal and with an energy density
significantly less than the kinetic energy of the ambient medium in the
frame as the projectile -- can significantly distort the projectile,
keeping it significantly more constrained in the plane perpendicular to
the ambient field, and keeping any stripped material inside the drape.

If a magnetic draping layer such as generated in these simulations were
astrophysically observable, it would be possible to get independent
measurements of the magnetic field strength in the ambient medium provided the
local gas density and the velocity of the moving object is known. The tools are
both the opening angle of the drape and the thickness of the magnetic field
layer -- but not through the layer's field strength.  Similarly, for a known
magnetic field strength, we would have an alternate measure of the velocity of
the projectile.

\bigskip 

\begin{acknowledgments}
  The authors thank M. Lyutikov for suggesting this work, Y. Lithwick
  for fruitful discussions, and M. Zingale, Y. Lithwick, M. Lyutikov,
  and the anonymous referee for helpful suggestions on this manuscript.  The authors gratefully acknowledge the financial support of
  the National Science and Engineering Research Council of Canada.
  The software used in this work was in part developed by the
  DOE-supported ASC / Alliance Center for Astrophysical Thermonuclear
  Flashes at the University of Chicago.  All computations were
  performed on CITA's McKenzie and Sunnyvale clusters which are funded
  by the Canada Foundation for Innovation, the Ontario Innovation
  Trust, and the Ontario Research Fund.  3D renderings were performed
  with OpenDX.  This work made use of NASA's Astrophysical Data
  System.
\end{acknowledgments}

\clearpage
\appendix
\section{DERIVATION OF THE MHD FLOW AROUND A MOVING BODY}
\label{sec:analytics}

\subsection{The exact MHD solution}
\label{sec:exact}

The full non-linear solution of the MHD flow around a moving and dynamically
evolving body is extremely complex because of its significant
three-dimensionality and non-linearity.  Owing to the range of scales involved
this problem is perfectly suited for an MHD adaptive mesh refinement
simulation.  In order to gain credibility and improve our understanding of the
properties of the numerical solution including its scaling behavior, we solve
the problem of an ideally conducting plasma around a spherical body
analytically.  To this end, we solve for the flow of a plasma with a frozen-in
magnetic field around a sphere to explore the characteristics of the magnetic
field near the surface of the body. We disregard any possible change in the
flow pattern by means of the back-reaction of the magnetic field. The same
problem has been investigated by \citet{1980Ge&Ae..19..671B} who find that the
energy density of the magnetic field forming in the wake behind the body
diverges logarithmically.  In passing by we correct the misconception that lead
to this unphysical behavior of their solution and derive a criterion for the
breakdown of our simplified analytical solution that we then successfully apply
to our numerical solution.

The governing equations of ideal MHD with infinity conductivity are given by
\begin{equation}
\label{eq:MHD}
\mbox{curl}(\vel\times\B) = \mathbf{0}
\quad\mbox{and}\quad
\mbox{div}\,\B = 0.
\end{equation}
We solve this system of equations outside the sphere for a given velocity field
that is derived for a viscous and incompressible flow around the sphere.
Without loss of generality, we choose the origin of our spherical coordinate
system to coincide with the center of the sphere with radius $R$ (Fig.~1) and
the $z$-axis being anti-parallel to the fluid velocity at infinity. The
potential flow solution of the pure hydrodynamical problem reads in spherical
coordinates as follows \citep{kotschin}:
\begin{equation}
\label{eq:v_solution}
\vel = \e_r \left(\frac{R^3}{r^3}-1\right) u\cos\theta +
     \e_\theta \left(\frac{R^3}{2 r^3}+1\right)u\sin\theta
   = -\u+\frac{R^3}{2 r^3}\,
   \left[3 \e_r (\u\cdot\e_r) - \u\right] ,
\end{equation}
where we employed the coordinate independent representation of the homogeneous
fluid velocity at infinity in the second step, $\u = \e_r (\u\cdot\e_r) +
\e_\theta (\u\cdot\e_\theta)$. Since for any stream line holds $\dd r/\v_r = r
\dd \theta/\v_\theta$, we can thus derive the equation of the line of flow
using Stoke's method of the stream function
\begin{equation}
\label{eq:stream}
p = r\sin\theta\sqrt{1 - \frac{R^3}{r^3}},
\end{equation}
where $p$ is the impact parameter of the given line of the flow from the
$z$-coordinate axis on an infinitely distant plane in the left half-space.  We
assume a homogeneous magnetic field at infinity in the left half-space
pointing towards the positive $y$-coordinate axis yielding the boundary
conditions for $\B$:
\begin{equation}
\label{eq:B_infinity}
\left.B_r\right|_\infty      = B_0 \sin\theta\sin\phi,\quad
\left.B_\theta\right|_\infty = B_0 \cos\theta\sin\phi,\quad
\left.B_\phi\right|_\infty   = B_0 \cos\phi.
\end{equation}
Writing Eqns.~(\ref{eq:MHD}) for the components yields
\begin{eqnarray}
  \label{eq:MHD:curl_r}
  \mbox{curl}_r(\vel\times\B):&&
  \partf{}{\theta}[\sin\theta(\v_r B_\theta-\v_\theta B_r)] + 
  \partf{}{\phi}(\v_r B_\phi) = 0,\\
  \label{eq:MHD:curl_t}
  \mbox{curl}_\theta(\vel\times\B):&&
  \partf{}{r}[r(\v_r B_\theta-\v_\theta B_r)] - 
  \frac{1}{\sin\theta}\partf{}{\phi}(\v_\theta B_\phi) = 0,\\
  \label{eq:MHD:curl_p}
  \mbox{curl}_\phi(\vel\times\B):&&
  \partf{}{r}(r \v_r B_\phi) + 
  \partf{}{\theta}(\v_\theta B_\phi) = 0,\\
  \label{eq:MHD:div}
  \mbox{div}\B: &&
  \frac{1}{r^2}\left[\partf{}{r}\left(r^2 B_r\right)\right] +
  \frac{1}{r\sin\theta}\left[\partf{}{\theta}(\sin\theta B_\theta)\right]+
  \frac{1}{r\sin\theta}\partf{B_\phi}{\phi} = 0.
\end{eqnarray}
By substituting (\ref{eq:v_solution}) into (\ref{eq:MHD:curl_p}) we obtain the
equation for $B_\phi$
\begin{equation}
  \label{eq:B_phi_part}
  \partf{}{r}B_\phi + \frac{\v_\theta}{r \v_r}\partf{}{\theta}B_\phi = 
  -\frac{3 B_\phi R^3}{2 r (r^3 - R^3)},
\end{equation}
where $\v_\theta/(r \v_r) = -\tan\theta\, (2 r^3+R^3) / [2r(r^3-R^3)]$.
Equation~(\ref{eq:B_phi_part}) is a linear inhomogeneous first-order partial
differential equation which can be solved by the method of characteristics. We
take $r$ as parameter in the characteristic equations and express the variables
$\theta$ and $\phi$ in terms of $r$, using
\begin{equation}
  \label{eq:characteristics}
  \frac{\dd B_\phi}{\dd r} = \partf{B_\phi}{r} + 
  \partf{B_\phi}{\theta}\partf{\theta}{t}\partf{t}{r} + 
  \partf{B_\phi}{\phi}\partf{\phi}{t}\partf{t}{r} = 
  \partf{B_\phi}{r} + \frac{\v_\theta}{r \v_r}\partf{B_\phi}{\theta}.
\end{equation}
Thus, on the line of the flow that is uniquely characterized by its impact
parameter $p$ at infinity, we obtain a first order ordinary differential
equation for $B_\phi$,
\begin{equation}
  \label{eq:B_phi_ord}
  \frac{\dd B_\phi}{\dd r} = -\frac{3 B_\phi R^3}{2 r (r^3 - R^3)}.
\end{equation}
Integrating this equation by the separation of variables yields the solution
for $B_\phi$ that contains a constant which is determined from the homogeneous
magnetic field at infinity (\ref{eq:B_infinity}),
\begin{equation}
  \label{eq:B_phi_sol}
 B_\phi = \frac{B_0 \cos\phi}{\dps\sqrt{1-\frac{R^3}{r^3}}}.
\end{equation}
To determine $B_r$ and $B_\theta$, we turn to Eqns.~(\ref{eq:MHD:curl_r}) and
(\ref{eq:MHD:curl_t}). By multiplying Eqn.~(\ref{eq:MHD:curl_r}) with $r$ and
(\ref{eq:MHD:curl_t}) with $\sin\theta$, defining $K\equiv r\sin\theta (\v_r
B_\theta - \v_\theta B_r)$, and combining (\ref{eq:MHD:curl_t}) and
(\ref{eq:MHD:curl_r}), we obtain the equation for $K$:
\begin{equation}
\label{eq:K_part}
\partf{K}{r} + \frac{\v_\theta}{r \v_r} \partf{K}{\theta} = 0.
\end{equation}
Equation~(\ref{eq:K_part}) can again be solved by the method of
characteristics as (\ref{eq:B_phi_part}) yielding $K=K_p$, where $K_p$ is a
constant on each flow line that is labeled with its impact parameter $p$.
Determining this constant from Eqn.~(\ref{eq:B_infinity}) and substituting for
$K$ and $K_p$ their values, we obtain the following equation that relates
$B_r$ and $B_\theta$,
\begin{equation}
\label{eq:B_t-B_r}
r\sin\theta\,(\v_r B_\theta - \v_\theta B_r) = 
- p u B_0 \sin\phi. 
\end{equation}
Substituting $B_\theta$, expressed in terms of $B_r$, from Eqn.~(\ref{eq:B_t-B_r})
and $B_\phi$ from Eqn.~(\ref{eq:B_phi_sol}) into the solenoidal condition for
$\B$ (\ref{eq:MHD:div}), we obtain the equation for $B_r$. Similarly,
substituting $B_r$, expressed in terms of $B_\theta$ and following the same
steps, leads to the equation for $B_\theta$:
\iftwocol
\begin{eqnarray}
\label{eq:B_r,t_part}
&&\partf{B_r}{r} + \frac{\v_\theta}{r \v_r}\partf{B_r}{\theta} + 
\left[\frac{2}{r} - \frac{2 r^3 + R^3}{2 r (r^3 - R^3)}
  \left(1 + \frac{1}{\cos^2\theta}\right)\right]\, B_r =
-\frac{B_0 \sin\phi \sin\theta}{r \sqrt{1 - \frac{R^3}{r^3}}\cos^2\theta}, 
\\
&&\partf{B_\theta}{r} + \frac{\v_\theta}{r \v_r}\partf{B_\theta}{\theta} + 
\left[\frac{2}{r} - \frac{2 r^3 + R^3}{2 r (r^3 - R^3)} + 
\frac{9 r^2 R^3}{(2 r^3 + R^3) (r^3 - R^3)}\right]\, B_\theta = 
 \frac{2 B_0\sin\phi\, (r^3 + 2 R^3)}
     {r \cos\theta\, (2 r^3 + R^3)\sqrt{1 - \frac{R^3}{r^3}}}.
\end{eqnarray}
\else
\begin{eqnarray}
\label{eq:B_r,t_part}
\lefteqn{
\partf{B_r}{r} + \frac{\v_\theta}{r \v_r}\partf{B_r}{\theta} + 
\left[\frac{2}{r} - \frac{2 r^3 + R^3}{2 r (r^3 - R^3)}
  \left(1 + \frac{1}{\cos^2\theta}\right)\right]\, B_r =
-\frac{B_0 \sin\phi \sin\theta}{r \sqrt{1 - \frac{R^3}{r^3}}\cos^2\theta}, } 
\hspace{0.5\hsize}\\
\lefteqn{
\partf{B_\theta}{r} + \frac{\v_\theta}{r \v_r}\partf{B_\theta}{\theta} + 
\left[\frac{2}{r} - \frac{2 r^3 + R^3}{2 r (r^3 - R^3)} + 
\frac{9 r^2 R^3}{(2 r^3 + R^3) (r^3 - R^3)}\right]\, B_\theta =} 
\hspace{0.5\hsize}\nonumber\\
&& \frac{2 B_0\sin\phi\, (r^3 + 2 R^3)}
     {r \cos\theta\, (2 r^3 + R^3)\sqrt{1 - \frac{R^3}{r^3}}}.
\end{eqnarray}
\fi
Both equations can again be solved by the method of characteristics,
expressing the variables $\theta$ and $\phi$ in terms of $r$ which we take to
be the independent parameter along the flow lines and using
Eqn.~(\ref{eq:stream}). Note that for a potential flow, the variable $\phi$ is
always constant on the line of the flow by symmetry. The resulting linear
inhomogeneous first-order ordinary differential equations are easily solved by
an integrating factor that is derived from the homogeneous equations, leading
to the solutions for $B_r$ and $B_\theta$,
\begin{eqnarray}
\label{eq:B_r_sol}
B_r &=& \frac{r^3 - R^3}{r^3}\cos\theta\,
\left[C_1\mp B_0\sin\phi \int_\xi^r\,\frac{p(r,\theta\,) r'^4\, \dd r'}
{\left(r'^3 - R^3 - p(r,\theta)^2 r'\right)^{3/2} \sqrt{r'^3 - R^3}}\,\right],\\
\label{eq:B_t_sol}
B_\theta &=& \frac{2 r^3 + R^3}{r^{5/2}\sqrt{r^3-R^3}}
\left[C_2 \pm 2 B_0 \sin\phi \int_\xi^r
\frac{r'^3\, (r'^3 + 2 R^3)\sqrt{r'^3-R^3}\,\dd r'}
{(2 r'^3+R^3)^2\sqrt{r'^3-R^3-p(r,\theta)^2\, r'}}\right],
\end{eqnarray}
where $C_1$ and $C_2$ are integration constants and $\xi$ is the initial value
for which $B_r$ and $B_\theta$ are known. The upper signs refer to the region
$0\leq \theta \leq \pi/2$, and the lower signs to $\pi/2\leq\theta \leq
\pi$. 

\subsection{The approximate MHD solution near the sphere}
\label{sec:approx}

We aim at understanding the behavior of the magnetic field in the region near
the sphere. To this end, we investigate the behavior of the integrals in
(\ref{eq:B_r_sol}) and (\ref{eq:B_t_sol}) for small impact parameters and keep
only the main terms with respect to $p$. We find that the integral in
(\ref{eq:B_r_sol}) diverges at the lower limit logarithmically for $\pi/2$ and
the integral in (\ref{eq:B_t_sol}) has a linear divergence at the lower limit.
Thus we will use (\ref{eq:B_r_sol}) in the region $0\leq \theta \leq \pi/2$
and (\ref{eq:B_t_sol}) in the region for $\pi/2\leq\theta \leq \pi$.

We divide the region of integration into two: the first from $\infty$ to $r_1$
where $r_1 > R$ is the radius of the sphere on which the asymptotic form of
the magnetic field changes, and the second from $r_1$ to $r_0$, where $r_0$ is
the radial value of the flow of line under consideration for $\theta = \pi/2$.
This implies that the following expansions only apply to small impact
parameters $p$ with $r_0 \leq r_1$.  By expanding the integrand of
(\ref{eq:B_r_sol}) in powers of $1/r$ for $r >r_1 > R$ and in the region
$0\leq \theta \leq \pi/2$, we determine $C_1 = 0$ and we recover the
homogeneous field at infinity with an accuracy to $\mathcal{O}(1/r)$. Near the
surface $r_1 > r > r_0$ we perform a change of the variable to $s=r-R$. We
define $s_1 = r_1-R$ and $s_0 = r_0-R$ and $s$ varies within $s_0< s <s_1$.
The equation of the line of flow (\ref{eq:stream}) has the form $p=\sqrt{3 s
  R}\,\sin\theta$ with an accuracy to $\mathcal{O}(s^{3/2})$ and from this we
obtain $s_0=p^2/(3\, R)$ for $\theta=\pi/2$ and $s=s_0$.

The value of $B_r$ in this region will be composed of two terms: the value of
the integral in (\ref{eq:B_r_sol}) from $\infty$ to $r_1$ with a factor to
leading order $\propto s^{3/2}$, and the value of the integral from $s_1$ to
$s$, which behaves like $\mathcal{O}(s^{1/2})$. Neglecting the first term in
comparison with the second, we obtain for $B_r$ with an accuracy to
$\mathcal{O}(s^{3/2})$ or $\mathcal{O}(p^3)$:
\begin{equation}
\label{eq:Br_Taylor}
B_r = -\frac{3s}{R} B_0\, p \sin\phi\cos\theta\,
\int_{s_1}^s\frac{s\,\dd s}{9\,(s^2 - s s_0)^{3/2}}.
\end{equation}
For impact parameters $p$ with $s_0 \leq s_1$, we obtain with an accuracy to
$\mathcal{O}(s_0/s_1)$:
\begin{equation}
\label{eq:B_r_s}
B_r = \frac{2}{3}B_0\sqrt{\frac{3s}{R}}\frac{\sin\theta}{1+\cos\theta}\sin\phi.
\end{equation}
Using Eqn.~(\ref{eq:B_t-B_r}) leads to the component $B_\theta$. Thus,
$B_\theta$ and $B_\phi$ near the sphere are determined by the formulae
\begin{eqnarray}
\label{eq:B_t_s}
B_\theta &=& B_0\sin\phi\,\sqrt{\frac{R}{3s}},\\
\label{eq:B_p_s}
B_\phi   &=& B_0\cos\phi\,\sqrt{\frac{R}{3s}}.
\end{eqnarray}
It turns out that these formulae are also correct for the region $\pi/2 \leq
\theta \leq \pi$ as follows from Eqn.~(\ref{eq:B_t_sol}). The integral in this
expression is regular for $s_0$, and by computing $B_\theta$ in the
approximation $s\ll R$ for $s=s_0$, we find $C_2 = \sin\phi\,R\,B_0/3$. Then
$B_\theta$ is equal to (\ref{eq:B_t_s}) with an accuracy to terms of order
$\mathcal{O}(s^{1/2})$. We obtain (\ref{eq:B_p_s}) by using
(\ref{eq:B_t-B_r}). Thus, Eqns.~(\ref{eq:B_r_s}) to (\ref{eq:B_p_s})
uniformly describe the field near the sphere with respect to the angle
$\theta$.

In order to facilitate comparison to our numerical solution, we transform the
approximate solution for $\B$ given by the components in the spherical
coordinate system (\ref{eq:B_r_s}) to (\ref{eq:B_p_s}) into Cartesian system
yielding
\begin{eqnarray}
\label{eq:B_xyz}
B_x &=& B_0\cos\phi\sin\phi\,(1-\cos\theta)\,\sqrt{\frac{R}{3s}}\,
        \left(\frac{2s}{R}-1\right),\\
B_y &=& B_0\sqrt{\frac{R}{3s}}\,\left[\sin^2\phi\,(1-\cos\theta)\,
        \left(\frac{2s}{R}-1\right) + 1\right] =
        B_x\, \tan\phi + B_0\sqrt{\frac{R}{3s}},\\
B_z &=& B_0\sin\phi\sin\theta\,\sqrt{\frac{R}{3s}}\,
\left(\frac{2s}{R}\,\frac{\cos\theta}{1+\cos\theta} - 1\right).
\end{eqnarray}
Note that we introduced the radial coordinate from the surface of the sphere
$s=r-R$ and that this solution applies only near the sphere with an accuracy
to $\mathcal{O}(s^{3/2})$ as well as for small impact parameters $p$ with an
accuracy to $\mathcal{O}(s_0/s_1)$.

Using the method of regularization of the integral in (\ref{eq:B_r_sol}) with
respect to the lower limit $\theta=\pi/2$, \citet{1980Ge&Ae..19..671B}
investigate the behavior of the magnetic field in the wake of the sphere.
They find that, when neglecting a term that scales as $\mathcal{O}(1/r)$,
$B_r$ is given by
\begin{equation}
\label{eq:B_r_wake}
B_r = \frac{4}{3} \frac{B_0\sin\phi}{p},
\end{equation}
which, with proximity to the axis of the wake $p\to 0$, leads to an unlimited
increase of $B_r \to \infty$. The magnetic lines of force that end at the
stagnation point are strongly elongated as the swipe around the sphere
parallel to the line of flow reaching from the stagnation point into the rear.
This leads to the unphysical increase of the magnetic field as it approaches
the symmetry line. While this might be the mathematically correct solution, it
leads to a logarithmic divergence of the energy density of the magnetic field
in the volume near the wall.

\del{
\section{VORTICITY GENERATION: BREAKDOWN OF THE SOLUTION}
\label{sec:vorticity}
The solutions for $\B$ in \S\ref{sec:analytics} disregard any
possible change in the flow pattern by means of the back-reaction of
the magnetic field. We thus expect the solution to be applicable until
the magnetic field becomes dynamically important in the wake of the
sphere as the draped magnetic field disconnects from the moving body
and influences the velocity field. We will derive a criterion for the
breakdown of our simplified analytical solution that assumed a
potential flow by considering the sources for vorticity generation.
The equation of motion for an inviscid and magnetized fluid without
gravity may be written in the form
\begin{equation}
\label{eq:Euler}
\frac{\dd \vel}{\dd t} = \partf{\vel}{t} + \vel\cdot\nabla\vel = 
-\frac{\nabla P}{\rho} + \vel \times \b,
\end{equation}
where we define the convective derivative in the first step and a scaled
magnetic field $\b = \B e / (m c)$.  Additionally, we have the continuity
equation and solenoidal condition for $\b$,
\begin{equation}
\label{eq:cont}
\partf{\rho}{t} + \nabla(\rho\vel) = 0
\quad\mbox{and}\quad
\nabla \b = 0.
\end{equation}
Applying the curl operator to Eqn.~(\ref{eq:Euler}) and identifying the
vorticity $\bomega = \nabla\times\vel$, we arrive at
\begin{equation}
\label{eq:step}
\frac{\dd }{\dd t}\left(\bomega + \b\right) = 
[(\bomega + \b)\cdot\nabla]\vel + \frac{\bomega + \b}{\rho}\frac{\dd \rho}{\dd t}+
\frac{1}{\rho^2}\nabla\rho\times\nabla P + \partf{\b}{t}.
\end{equation}
This may rewritten yielding
\begin{equation}
\label{eq:flux_freezing}
\frac{\dd }{\dd t}\left(\frac{\bomega + \b}{\rho}\right) = 
\left[\left(\frac{\bomega + \b}{\rho}\right)\cdot\nabla\right]\vel + 
\frac{1}{\rho^3}\nabla\rho\times\nabla P + \frac{1}{\rho}\partf{\b}{t}.
\end{equation}
Neglecting the last two terms, this equation describes the condition that the
sum of the magnetic field lines and the vorticity are `frozen' in the
plasma. This can be seen by considering the evolution of an infinitesimal vector
$\delta \x$ connecting two neighboring fluid parcels, as the fluid moves with
the velocity field. The point initially at position $\x$ at time $t$ will be
displaced to the position $\x+\vel(\x)\Delta t$ at time $t+\Delta t$. The
neighboring point initially at $\x+\delta\x$ at time $t$ will be displaced to
the position $\x+\delta\x+\vel(\x+\delta\x)\Delta t$ at time $t+\Delta t$. Hence
this `frozen' connecting line evolves according to 
\begin{equation}
\label{eq:x_flux_freezing}
\frac{\dd }{\dd t}\left(\delta\x\right) = 
(\delta\x\cdot\nabla)\vel,
\end{equation}
which resembles Eqn.~(\ref{eq:flux_freezing}) if we neglect the last two terms
and identify $\delta\x = \eps (\bomega + \b)/\rho$ initially, where $\eps>0$
is a small quantity. Since the differential equation is true for any time, the
same relation will hold for all times for the sum of the magnetic field lines
and the vorticity. We draw the following conclusions from
Eqn.~(\ref{eq:flux_freezing}). (i) Magnetic field lines and vorticity are
frozen into the flow if $\nabla\rho$ is parallel to $\nabla P$ and at the same
time $\partial \B / (\partial t) = 0$. (ii) Vorticity (and magnetic fields)
are generated if $\nabla\rho$ is not aligned with $\nabla P$ or if $\partial
\B / (\partial t) \ne 0$. (iii) In the rest frame of the moving fluid, the
magnetic draping layer and the magnetic shoulder past the bullet represents a
change of the magnetic field, where $\partial \B / (\partial t) =
\nabla\times(\vel\times\B)\ne 0$, assuming infinite conductivity. Thus, as the
fluid enters the draping layer or moves across this shoulder, vorticity is
generated. The resulting velocity field can not any more described by the
potential flow solution which causes the analytical solution to break down at
the magnetic draping layer and behind the magnetic shoulder.
}

\clearpage

\bibliographystyle{plainnat}
\bibliography{magneticdraping}

\clearpage

\end{document}